\documentclass[bibyear]{aa} 

\usepackage{graphicx}
\usepackage{txfonts}
\usepackage{natbib}
\usepackage{bm}
\usepackage{multirow}
\usepackage{multicol}
\usepackage[dvipsnames]{xcolor}
\usepackage{dirtytalk}

\usepackage{listings}

\newcommand{\mh}{[{\rm M}/{\rm H}]}
\newcommand{\feh}{$\left[\rm{Fe}/\rm{H}\right]$}
\newcommand{\afe}{$\left[\alpha/\rm{Fe}\right]$}

\newcommand{\msun}{{\rm M}_\odot}

\newcommand{\g}{{\it Gaia}}

\newcommand{\gedr}{{\it Gaia} EDR3}
\newcommand{\jdr}{J-PLUS DR3}

\newcommand{\rsds}{r}
\newcommand{\mrsds}{m_{r} }

\newcommand{\dm}{ f }
\newcommand{\jpxg}{JP$\times$G }
\newcommand{\jpxgk}{JP$\times$G$_{\rm 5 kpc}$ }
\newcommand{\jpxgf}{JP$\times$G$_{\rm 5kpc}^{r16.5}$ }

\usepackage[colorlinks=true, linkcolor=blue, citecolor=blue, urlcolor=blue]{hyperref}
\begin{document}

\title{J-PLUS: Reconstructing the Milky Way Disc’s star formation history with twelve-filter photometry}

\subtitle{}

\author{ J. A. Alzate-Trujillo\inst{\ref{CEFCA}}
    \and A. del Pino\inst{\ref{IAA}}\inst{\ref{CEFCA}}
    \and C.~L\'opez-Sanjuan\inst{\ref{CEFCA},\ref{UA}}
    \and A. Hidalgo\inst{\ref{CEFCA}}
    \and S. Turrado-Prieto\inst{\ref{CAB}}
    \and Vinicius Placco\inst{\ref{NSF}}
    \and Paula Coelho\inst{\ref{USP}}
    \and Haibo Yuan\inst{\ref{BNU}}
    \and Luis Lomel\'i-N\'u\~nez\inst{\ref{VO}}
    \and Gustavo Bruzual\inst{\ref{IRYA}}
    \and F. Jiménez-Esteban\inst{\ref{CAB}}
    \and Eduardo Telles\inst{\ref{ON}}
    \and Borja Anguiano\inst{\ref{CEFCA}}
    \and Alvaro Alvarez-Candal\inst{\ref{IAA}}
    \and A.~J.~Cenarro\inst{\ref{CEFCA},\ref{UA}}
    \and D.~Crist\'obal-Hornillos\inst{\ref{CEFCA}}
    \and C.~Hern\'andez-Monteagudo\inst{\ref{IAC},\ref{ULL}}
    \and A.~Mar\'{\i}n-Franch\inst{\ref{CEFCA},\ref{UA}}
    \and M.~Moles\inst{\ref{CEFCA}}
    \and J.~Varela\inst{\ref{CEFCA}}
    \and H.~V\'azquez Rami\'o\inst{\ref{CEFCA},\ref{UA}}
    \and J.~Alcaniz\inst{\ref{ON}}
    \and R.~A.~Dupke\inst{\ref{ON},\ref{MU}}
    \and A.~Ederoclite\inst{\ref{CEFCA},\ref{UA}}
    \and L.~Sodr\'e Jr.\inst{\ref{USP}}
    \and R.~E.~Angulo\inst{\ref{DIPC},\ref{ikerbasque}}
}

\institute{
Centro de Estudios de F\'{\i}sica del Cosmos de Arag\'on (CEFCA), Plaza San Juan 1, 44001 Teruel, Spain\label{CEFCA}
\and Instituto de Astrofísica de Andalucía – Consejo Superior de Investigaciones Científicas (IAA-CSIC), Glorieta de la Astronomía S/N,E-18008, Granada, Spain\label{IAA}
\and Centro de Astrobiolog\'{\i}a (CAB), CSIC-INTA, Camino Bajo del Castillo s/n, E-28692, Villanueva de la Ca\~{n}ada, Madrid, Spain.\label{CAB}
\and NSF NOIRLab, Tucson, AZ 85719, USA\label{NSF}
\and Department of Astronomy, Beijing Normal University, Beijing, 100875, People's Republic of China\label{BNU}
\and Universidade Federal do Rio de Janeiro, Observat\'orio do Valongo, Ladeira do Pedro Ant\^onio, 43, Sa\'ude CEP 20080-090 Rio de Janeiro, RJ, Brazil\label{VO}
\and Instituto de Radioastronom\'ia y Astrof\'isica, Universidad Nacional Aut\'onoma de M\'exico, Morelia, Michoac\'an 58089, M\'exico\label{IRYA}
\and Unidad Asociada CEFCA-IAA, CEFCA, Unidad Asociada al CSIC por el IAA y el IFCA, Plaza San Juan 1, 44001 Teruel, Spain\label{UA}
\and Instituto de Astrof\'{\i}sica de Canarias, La Laguna, 38205, Tenerife, Spain\label{IAC}
\and Departamento de Astrof\'{\i}sica, Universidad de La Laguna, 38206, Tenerife, Spain\label{ULL}
\and Observat\'orio Nacional - MCTI (ON), Rua Gal. Jos\'e Cristino 77, S\~ao Crist\'ov\~ao, 20921-400 Rio de Janeiro, Brazil\label{ON}
\and University of Michigan, Department of Astronomy, 1085 South University Ave., Ann Arbor, MI 48109, USA\label{MU}
\and Instituto de Astronomia, Geof\'{\i}sica e Ci\^encias Atmosf\'ericas, Universidade de S\~ao Paulo, 05508-090 S\~ao Paulo, Brazil\label{USP}
\and Donostia International Physics Centre (DIPC), Paseo Manuel de Lardizabal 4, 20018 Donostia-San Sebastián, Spain\label{DIPC}
\and IKERBASQUE, Basque Foundation for Science, 48013, Bilbao, Spain\label{ikerbasque}
}

\date{Received XXXX, 2025; XXXXX, XXXX}

\abstract
{Wide-field, multi-filter photometric surveys enable the reconstruction of the Milky Way’s star formation history (SFH) on Galactic scales and provide complementary insights into disc assembly. The twelve-filter system of the Javalambre Photometric Local Universe Survey (J-PLUS) is particularly suitable, as its colours trace stellar chemical abundances and help mitigate the age–metallicity degeneracy in colour–magnitude diagram fitting.}
{We aim to recover the SFH of the Milky Way disc and separate its chemically distinct components by combining J-PLUS DR3 photometry with Gaia astrometry. We also test the potential of isochrone fitting to estimate ages and metallicities for individual stars as proxies for disc evolutionary trends.}
{We fit magnitudes and parallaxes of $1.38\times10^{6}$ stars using a Bayesian multiple-isochrone technique. The bright region of the colour–absolute-magnitude diagram ($M_{r}\le4.2$ mag) constrains stellar ages, while the faint region provides an empirical metallicity prior that mitigates the age–metallicity degeneracy. Both PARSEC and BaSTI isochrones, in solar-scaled and $\alpha$-enhanced versions, are adopted.}
{The recovered SFH shows two sequences: an $\alpha$-enhanced population forming rapidly between $12.5$ and $8$ Gyr ago, enriching from $\mathrm{[M/H]}\sim-0.6$ to $0.1$ dex; and a solar-scaled sequence emerging $\sim8$ Gyr ago, dominating after $\sim7$ Gyr with slower enrichment, reaching solar metallicity by $3$ Gyr. Metal-rich ($\mathrm{[M/H]}\gtrsim0$) stars are confined to $|z_{\mathrm{GC}}|\lesssim1$ kpc, whereas metal-poor ($\mathrm{[M/H]}\lesssim-0.5$) stars reach $|z_{\mathrm{GC}}|\sim2$ kpc.}
{Simultaneous fitting of solar-scaled and $\alpha$-enhanced isochrones reveals distinct formation epochs for the thin and thick discs. J-PLUS multi-filter photometry, combined with Gaia parallaxes, effectively mitigates age–metallicity degeneracies and enables detailed mapping of the Milky Way’s temporal and chemical evolution.}


   \maketitle
%

\section{Introduction}\label{sec:intro}

The study of the Milky Way (MW) is experiencing significant advances, fueled by an unprecedented convergence of astrometric, spectroscopic, and photometric data. On the astrometric front, the European Space Agency’s \textit{Gaia} mission \citep{prusti16,brown18,vallenari23} has provided precise parallaxes and proper motions for over a billion stars, delivering a refined view of our Galaxy’s structure and kinematics \citep{brown21}. Meanwhile, ground-based spectroscopic surveys, including Sloan Extension for Galactic Understanding and Exploration \citep[SEGUE;][]{yanny09}, Large Sky Area Multi-Object Fiber Spectroscopy Telescope \citep[LAMOST;][]{luo15, Yan22}, Radial Velocity Experiment \citep[RAVE;][]{steinmetz20a, steinmetz20b}, the Galactic Archaeology with HERMES \citep[GALAH;][]{deSilva15,buder21}, Apache Point Observatory Galactic Evolution Experiment \citep[APOGEE;][]{majewski17}, WHT Enhanced Area Velocity Explorer \citep[WEAVE;][]{dalton16,jin224}, 4-metre Multi-Object Spectroscopic Telescope \citep[4MOST;][]{deJong19}, and Dark Energy Spectroscopic Instrument \citep[DESI;][]{cooper23}, complement \textit{Gaia} with radial velocities and elemental abundances.
In addition, asteroseismology missions such as \textit{Kepler} \citep{borucki10}, \textit{K2} \citep{howell14} and \textit{TESS} \citep{ricker15} show how stellar oscillations can constrain ages,
especially for evolved stars \citep[e.g.,][]{chaplin20}.

One goal of these combined efforts is to reconstruct the Milky Way’s Star Formation History (SFH), defined as the temporal evolution of its star formation rate and associated chemical enrichment.
The SFH encodes the interplay between gas accretion, stellar evolution, and dynamical processes over billions of years, allowing us to identify events that shaped our Galaxy, assess the impact of mergers and interactions on its structure, and place the MW in the broader context of galaxy formation.
The MW is commonly divided into thin disc, thick disc, halo, and bulge, each hosting stars with distinct ages and metallicities that reflect specific formation epochs and evolutionary paths \citep[][and references therein]{bland_h16}.
The thin disc is typically younger and metal-rich, indicative of ongoing star formation and enrichment \citep{lian20,gallart24}; the thick disc comprises older, more metal-poor stars, likely formed during a turbulent early epoch influenced by mergers or dynamical heating \citep{sharma19,xiang22}; and the halo, which houses some of the oldest and most metal-poor stars \citep{savino20,kim25}, informs the initial assembly of the Galaxy, including accretion of satellites.

Disentangling the formation and evolution of these components requires detailed ages and metallicities on Galactic scales, a demanding task because ages are not directly observed but inferred by comparing stellar properties to stellar evolution models \citep{soderblom10}.
Various methods address this, each with distinct strengths and limitations.
Isochrone fitting with spectroscopic parameters remains widely used for individual ages.
By comparing effective temperature, surface gravity, and metallicity (and/or broadband photometry) with model grids, this approach estimates ages for extensive samples.
\citet{kordopatis23} fit ages using \g~DR3 spectroscopic and photometric data, obtaining relative uncertainties below 50\%, primarily for main-sequence turn-off (MSTO) stars.
Large spectroscopic surveys, combined with \g, have been crucial: for example, \citet{queiroz23} derived ages with typical relative uncertainties of $\sim$30\% for MSTO and subgiant branch (SGB) stars, and around 15\% for SGB stars alone.
Spectroscopic programs, however, face complex selection functions. Asteroseismology offers a different route for evolved stars, using oscillation frequencies to constrain mass and evolutionary state \citep{mathur12,chaplin20}.
While it can yield $\lesssim$10\% age uncertainties \citep{grossman25}, current samples are smaller and brightness limited, making it less feasible for large-scale Galaxy studies.

Colour-Magnitude Diagram (CMD) fitting of entire stellar populations offers a complementary, more global perspective on the SFH. Rather than assigning ages star-by-star, CMD fitting compares observed diagrams with synthetic ones constructed from stellar evolution models, thereby constraining the temporal and chemical evolution of whole populations \citep{tosi91,gallart99,aparicio04,dolphin02,hidalgo09}. This methodology has been highly successful in the Local Group, where deep CMDs based on broadband photometry have delivered detailed SFHs for nearby galaxies \citep{cole07,monelli10,delpino13,weisz14,delpino17,ruizlara20b}. Until recently, its application to the MW was limited by distance accuracy, but \g’s precise parallaxes now enable colour–absolute magnitude diagram (CAMD) analyses of Galactic populations out to several kiloparsecs \citep{bernard18,alzate21,daltio21,gallart19, mazzi24}, allowing more comprehensive reconstructions of the MW’s SFH.
Such three-dimensional analyses require accurate dust maps to correct the stellar photometry for extinction, otherwise, the inferred SFH may be biased and the identification of multiple stellar populations compromised.

Using this approach, \citet{ruizlara20} showed that the MW’s SFH was strongly influenced by the Sagittarius dwarf spheroidal, with successive star formation peaks coinciding with its pericentre passages. More recently, \citet{gallart24} obtained the SFH for stars within 100~pc of the Sun \citep{smart21}, revealing coeval populations with different metallicities that challenge conventional enrichment in the Solar Neighborhood, suggesting radial migration or a dramatic event. Applying the same CAMD fitting methodology, \citet{alvar25} studied the thin and thick discs via kinematically selected stars in a cylindrical region of $250$pc radius and 1kpc height ($-500<z/{\rm pc}<500$) centered on the Sun. They found a thick disc dominated by stars older than $10$~Gyr with rapid chemical enrichment and suggested that the transition to the thick disc may be linked to the accretion of the \g-Sausage Enceladus \citep[GSE;][]{belokurov18,helmi18} system. The GSE itself is analyzed in \citet{koda25} using the same technique, CAMD fitting over kinematically selected samples, to derive its SFH.

Thus far, studies based on flux-limited photometry provide among the most complete Galaxy SFH views, being less subject to the selection biases and small samples inherent to spectroscopic or asteroseismic data. However, broadband photometry alone yields coarser stellar parameter constraints and stronger age–metallicity degeneracies.

In recent years, multi-filter photometry has emerged as a particularly effective balance of precision and efficiency. By sampling specific wavelength ranges and absorption features, such surveys enable reliable metallicity and abundance measurements without the demands of high-resolution spectroscopy. The Javalambre-Photometric Local Universe Survey \citep[J-PLUS;][]{cenarro19} exemplifies this new generation, employing twelve filters (including seven narrow bands) that capture key diagnostic lines such as H$\alpha$ and Ca II H and K \citep[e.g.,][]{whitten19, huang24}. Its third data release (DR3) provides photometry for over 47 million sources across $\sim$3,000 deg$^2$ \citep{sanjuan24}, delivering a homogeneous dataset on the AB system. Integrating {\g} astrometry with J-PLUS photometry allows CAMDs spanning the thin and thick discs and reaching into the halo over large volumes, offering a panoramic view of stellar populations with diverse ages and metallicities. Moreover, the synergy of multi-filter photometry improves parameter estimates relative to broad-passband data, reducing biases and enabling more robust SFH determinations.

In this paper, we revisit the MW's SFH using a robust approach that combines J-PLUS DR3 with \g.
Our method builds on the multiple isochrone fitting scheme of \citet{small13} and the Bayesian framework introduced by \citet{alzate21}.
Each star’s photometry is represented as an $n$-dimensional point in magnitude space (one dimension per filter),
with observations and uncertainties modeled as the product of $n$ independent Gaussian distributions.
This setup simplifies adding filters and is especially advantageous in an era when multi-filter surveys such as J-PLUS, the Javalambre Physics of the Accelerating Universe Astrophysical Survey \citep[J-PAS;][]{bonoli21},
the Southern Photometric Local Universe Survey \citep[S-PLUS;][]{oliveira19}, and forthcoming wide-field projects like the Legacy Survey of Space and Time \citep[LSST;][]{ivezic19} will greatly expand stellar population studies.
Overall, this versatile technique uses all available photometry to constrain stellar ages, metallicities, and, ultimately, the Galaxy’s SFH.

Throughout this paper, we focus on reconstructing the SFH of a J-PLUS sample of the Milky Way disc to provide insights into formation episodes and timescales, star formation bursts, and chemical enrichment. By comparing our findings with previous SFH studies in both the local solar vicinity and on larger scales, we demonstrate the consistency of our results and place them in the context of thick and thin discs formation and evolution, further underscoring the advantages of multi-filter photometric surveys for Galactic archaeology.


\section{Data}\label{sec:data}
Conducted at the Observatorio Astrofísico de Javalambre (OAJ; Teruel, Spain; \citealt{oaj}) with the 83 cm Javalambre Auxiliary Survey Telescope (JAST80) and the $9200 \times 9200$ pixels camera T80Cam \citep{t80cam},
J-PLUS has observed millions of celestial objects, including stars, galaxies, and quasars, with a 12-band photometric system \citep{cenarro19}.
Table~\ref{tab:filters} shows the central wavelengths and widths of the filter set.
Specifically, four broad filters ($u, g, r, i, z$) and seven narrow filters ($J0378, J0395, J0410, J0430, J0515, J0660$ and $J0861$), each designed to detect specific spectral features ($[{\rm O}_{\rm II}]$, Ca H+K, H$\delta$, G-band, Mg b triplet, H$\alpha$ and Ca triplet, respectively).
These emission/absorption lines provide valuable data for studying characteristic properties of Milky Way stellar populations, galaxy formation and evolution, and the large-scale structure of the universe.

The J-PLUS DR3, published in July 2022, is a catalog that contains observations collected between November 2015 and February 2022. {\jdr} covers approximately 3192 square degrees of the sky, (with about 2881 square degrees after applying masks to remove contaminated regions). The release provides measurements of about 47.4 million astronomical objects, of which nearly 29.8 million have brightness values of $\mrsds\leq$21, making it a relevant resource for analysing both nearby and distant objects.

\begin{table}[ht]
 \caption{\label{tab:filters}J-PLUS filter set. The first column ($k$) denotes the index number assigned to each filter. In Section~\ref{sec:BayesInfer}, we define the absolute $J^{k}$ magnitude, where each component ($J^{1}, J^{2}, \ldots, J^{12}$) corresponds to the filters listed in this table. This indexing facilitates the reference and manipulation of filter-specific data within our analysis.}
 \centering
 \begin{tabular}{cccc}
 \hline
 $k$     &    Filter    &    Wavelength (\AA)    &    FWHM(\AA)     \\
 \hline
 1   &      $u$       &          3485          &      508        \\
 2   &      $g$       &          4803          &      1409       \\
 3   &      $r$       &          6254          &      1388       \\
 4   &      $i$       &          7668          &      1535       \\
 5   &      $z$       &          9114          &      1409       \\
 \hline
 6   &    $J0378$     &          3785          &      168        \\
 7   &    $J0395$     &          3950          &      100        \\
 8   &    $J0410$     &          4100          &      200        \\
 9   &    $J0430$     &          4300          &      200        \\
 10   &    $J0515$     &          5150          &      200        \\
 11   &    $J0660$     &          6600          &      138        \\
 12   &    $J0861$     &          8610          &      400        \\
 \hline
 \end{tabular}
\end{table}

In order to obtain a J-PLUS sample representative of the stellar populations in the Milky Way, we selected objects from the {\jdr} catalog with good-quality detections (\texttt{FLAGS}$<3$, \texttt{MASK\_FLAGS}$<1$, and \texttt{NORM\_WMAP\_VAL}$>0.8$ across all filters). Full details are provided in Appendix~\ref{app:selfun}. Following this selection, the final sample contains 6,662,359 stars.

We obtained the $3$ arcsec diameter aperture photometry and its associated uncertainties from the
\texttt{MagABDualPointSources}\footnote{All the {\jdr} catalogs are available at
\url{https://archive.cefca.es/catalogues/jplus-dr3/help_adql.html}} catalog for each object and filter.
The magnitude values were corrected for flux loss due to the aperture limit and for contamination from nearby sources.
Sources affected by border effects or other CCD biases were filtered using \texttt{SExtractor} flags.

To refine our stellar sample, we used the probabilistic classification provided by the Bayesian artificial neural networks to classify J-PLUS objects \citep[BANNJOS;][]{delpino24}, which assigns each source a full posterior probability of being a star, galaxy, or QSO. Following the selection criteria of \citet{delpino24}, we retained only those objects whose stellar probability exceeds the 2$\sigma$ threshold. This conservative cut substantially screened out non-stellar contaminants while preserving the overall completeness of the sample.

We cross-matched the refined stellar sample with {\g} Early Data Release 3 (\gedr) using the \texttt{xmatch\_gaia\_edr3} catalog.
We note that, as {\g} DR3 adopts the same astrometric solution as \gedr \citep{vallenari23}, the results presented here remain fully consistent with Gaia DR3.
From \g, we extracted parallaxes ($\varpi$), parallax uncertainties ($e_{\varpi}$), and the Renormalised Unit Weight Error (RUWE).
To ensure reliable distance measurements, we restricted our analysis to stars with $e_{\varpi}/\varpi\leq0.3$ and RUWE $< 1.4$, which filters out sources with potentially spurious parallaxes.
We adopt 5 kpc as the distance limit beyond which the stellar counts from the disc become negligible.
This choice also provides a good compromise with the astrometric and photometric uncertainties within this range, given the completeness limits described in the following paragraphs.
The resulting cross-matched catalogs are referred to as \jpxg ($6,122,492$ stars) and \jpxgk ($3,966,815$ stars).

The J-PLUS DR3 sky is divided into tiles, each covering 2 square degrees observed with all twelve filters. Within each tile, the pipeline employs $\rsds$-band images for source detection. Tile completeness is assessed using a put-and-recover process, where synthetic sources are randomly generated and recovered through the same source extraction applied to real data. The 50\% detection magnitude for point-like sources, \texttt{M50S}, is reported in the \texttt{TileImage} catalog.

\begin{figure}[ht]
   \centering
   \includegraphics[width=0.95\columnwidth]{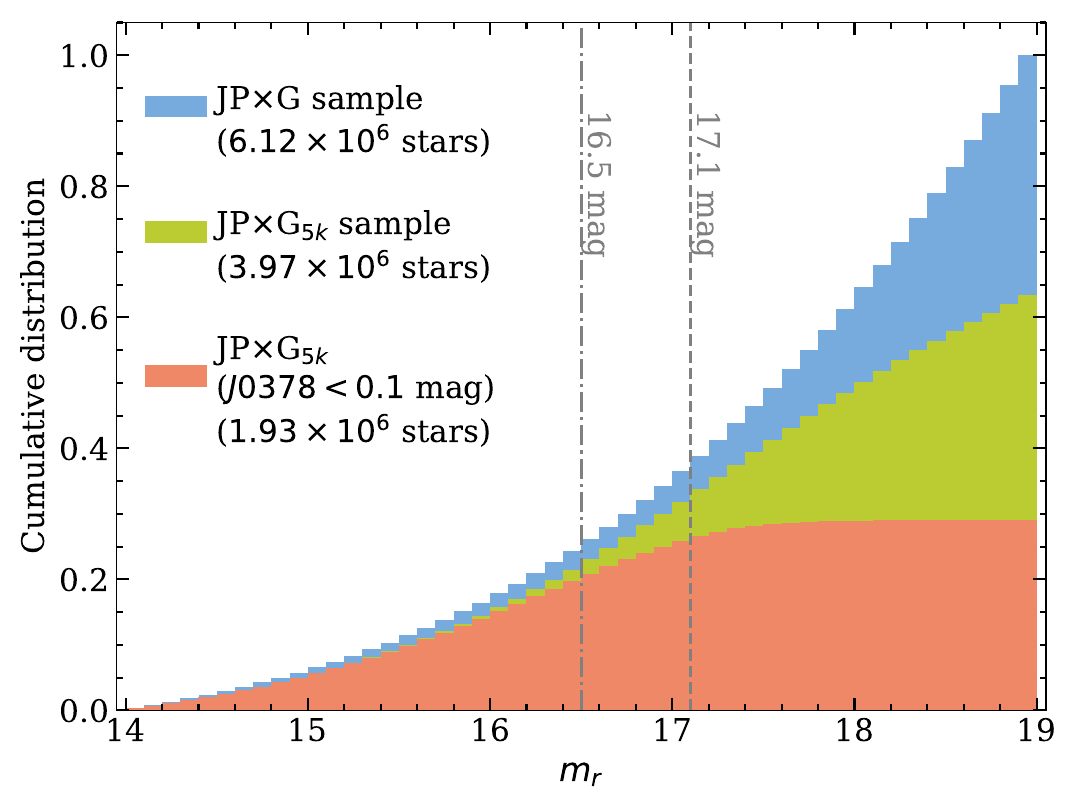}
   \caption{Cumulative counts of the \jpxg, \jpxgk and \jpxgk($J0378<0.1$ mag) samples. The dash-dotted and dashed lines point out the limits $\rsds=16.5$ mag and $\rsds=17.1$ mag, where the \jpxgk and \jpxgk($J0378<0.1$ mag) counts drops $10\%$ below the \jpxg counts.}
   \label{fig:cumul_rSDSS}
\end{figure}

However, after applying the selection criteria described in Appendix \ref{app:selfun}, the cross-match with \gedr, and the astrometric filters, the \texttt{M50S} limit is no longer valid. To address this, we empirically estimated the completeness of the \jpxgk sample using the \jpxg sample as a fiducial reference. The \jpxg sample maintains a completeness level above 90\% for sources with $\mrsds \leq 19$ mag. For the \jpxgk sample, the cumulative stellar counts drop by 10\% relative to \jpxg at $\mrsds = 17.1$ mag (see Fig.~\ref{fig:cumul_rSDSS}). To mitigate uncertainties associated with the $J0378$ fluxes, which have the largest errors among the J-PLUS filters, we imposed an additional restriction of $e_{J0378} \leq 0.1$ mag ($1,925,850$ stars). Under this condition, the stellar counts drop by 10\% at $\mrsds = 16.5$ mag.
Table~\ref{tab:stats} shows the percentiles of the distance modulus and photometry errors.

To ensure robust results, we applied bright and faint magnitude limits to avoid incompleteness modeling outside the reliable magnitude range.
Stars brighter than $\mrsds = 14.0$ mag were excluded, as they saturate the CCDs of the J-PLUS camera.
Adding this to the completeness considerations described in the previous paragraph, we estimate that the \jpxgk\ sample is, at least, $\approx 80\%$ complete between $\mrsds = 14.0$~mag (bright limit) and $\mrsds = 16.5$~mag (faint limit).

We corrected the photometry for dust attenuation using the extinction map Bayestar from \cite{green15,green18}\footnote{\url{https://dustmaps.readthedocs.io/en/latest/index.html}} and the $R_{V}$ values in Table 6 of \cite{schafly11}. The extinction curve from \cite{cardelli89} was used to derive extinction corrections for each J-PLUS filter. These calculations were performed with the Python package \texttt{extinction}\footnote{\url{https://extinction.readthedocs.io/en/latest/}}.

\begin{figure*}[ht]
\centering
\includegraphics[width=1.0\textwidth]{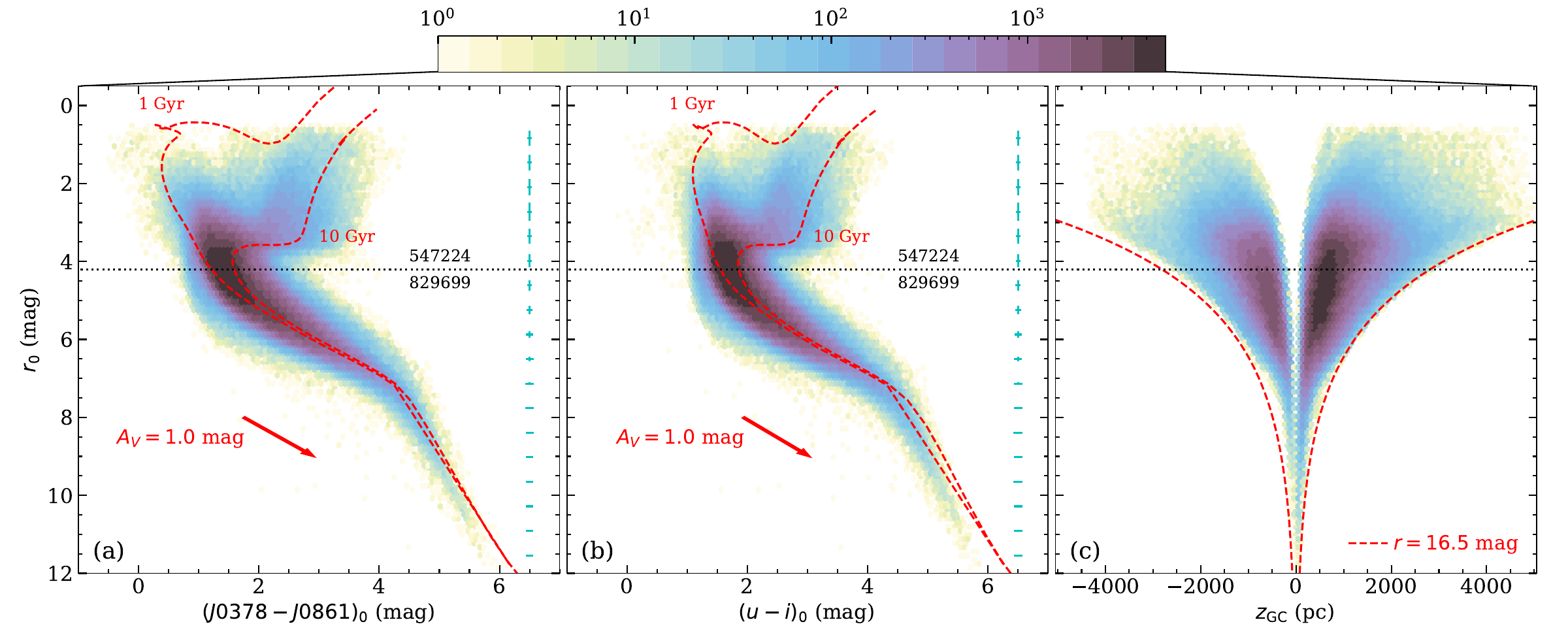}
\caption{Hess diagram of the \jpxgf sample for a) $J0378-J0861$ and b) $u-i$. Panel c) shows the distribution of stars in the ($\rsds$ vs $z_{GC}$) plane. The stellar colours and magnitudes are corrected by the effect of dust \citep{green18, schafly11}. The {\it dotted} horizontal line indicates the absolute $\rsds$ magnitude  equals to 4.2 mag, separating the regions we used to extract the information of ages and metallicities. The {\it arrow} illustrates the dimming and reddening vector corresponding to $A_{V}=1$ mag. The {\it crosses} arranged vertically show the colour and magnitude average errors, computed for all star within interval of size $\Delta \rsds = 1$ mag.}
\label{fig:jplus_cmd}
\end{figure*}

The resulting sample, which we call \jpxgf (1,376,923 stars), forms the basis for our analysis. Figure~\ref{fig:jplus_cmd} shows the CAMD for this sample, using the $(J0378-J0861)$ and $(u-i)$ colours in panels (a) and (b), respectively. The dotted line divides the CAMD into two regions (Sec.~\ref{ssec:samp_cmd}):  
\begin{itemize}
    \item Bright region ($\rsds \leq 4.2$ mag)\footnote{For simplicity, and hereafter, we denote the absolute $\rsds$ magnitude in the CAMD simply as $\rsds$.}: sensitive to both stellar age and metallicity.  
    \item Faint region ($\rsds > 4.2$ mag): primarily informs on chemical abundances.  
\end{itemize}

The red arrow in Fig.~\ref{fig:jplus_cmd} indicates the extinction vector for $A_V = 1$ mag.

\begin{table}[ht]
\renewcommand{\arraystretch}{1.25} 
\begin{center}
 \caption{Error percentiles of the photometric magnitudes contained in the \jpxgk ($J0378<0.1$ mag) sample, the distance modulus (DM) and the visual extinction. The percentiles were computed for two cases, stars bringter than 16.5 mag and  17.0 mag.}
 \label{tab:stats}
 \resizebox{9cm}{!}{
 \begin{tabular}{cccccccc}
 \hline
 \multirow{2}{*}{}   &  \multicolumn{3}{c}{$\mrsds\leq16.5$ mag}  &   &  \multicolumn{3}{c}{$\mrsds\leq17.1$ mag}      \\
                     & $p_{10}$ & $p_{50}$ & $p_{90}$ &   & $p_{10}$ & $p_{50}$ &  $p_{90}$  \\ \cline{2-4} \cline{6-8}

  $e_{\rm DM}$ & 0.03893  & 0.11233  & 0.27819  &   & 0.04278  & 0.13795  & 0.36848    \\
  $e_{\rm A_{V}}$ & 0.00700  & 0.01390  & 0.02480  &   & 0.00700  & 0.01400  & 0.02540 \\ \cline{2-4} \cline{6-8}
  
  $e_{u}$      & 0.02431  & 0.04136  & 0.07133  &   & 0.02735  & 0.05532  & 0.20285 \\
  $e_{g}$      & 0.01065  & 0.01804  & 0.03035  &   & 0.01179  & 0.02004  & 0.03281 \\
  $e_{r}$      & 0.00742  & 0.01286  & 0.02198  &   & 0.00809  & 0.01364  & 0.02293 \\
  $e_{i}$      & 0.00614  & 0.01025  & 0.01694  &   & 0.00669  & 0.01105  & 0.01808 \\
  $e_{z}$      & 0.00579  & 0.00929  & 0.01394  &   & 0.00632  & 0.01060  & 0.01643 \\ \cline{2-4} \cline{6-8}

  $e_{J0378}$  & 0.02297 & 0.03903 & 0.06957 &  & 0.02581 & 0.05175 & 0.18991\\
  $e_{J0395}$  & 0.02212 & 0.03694 & 0.06154 &  & 0.02479 & 0.04834 & 0.15106\\
  $e_{J0410}$  & 0.01691 & 0.02721 & 0.04158 &  & 0.01889 & 0.03325 & 0.06621\\
  $e_{J0430}$  & 0.01587 & 0.02538 & 0.03879 &  & 0.01770 & 0.03080 & 0.05618\\
  $e_{J0515}$  & 0.01212 & 0.01913 & 0.02960 &  & 0.01348 & 0.02262 & 0.03614\\
  $e_{J0660}$  & 0.00888 & 0.01400 & 0.02191 &  & 0.00977 & 0.01575 & 0.02380\\
  $e_{J0861}$  & 0.00705 & 0.01119 & 0.01627 &  & 0.00770 & 0.01286 & 0.01969\\
 \hline
 \end{tabular}
 }
\end{center}
\end{table}

\section{Modeling the colour-absolute-magnitude diagram}\label{sec:method}

CAMD fitting with stellar isochrones is a widely used technique for inferring the ages of resolved stellar populations (see Sec.~\ref{sec:intro}). The simplest case involves determining the age of a population whose CAMD closely matches that of a theoretical single stellar population, meaning all stars share the same age and metallicity. However, the Galactic disc is far more diverse, encompassing stars that formed over millions to billions of years and exhibiting significant chemical variations. Consequently, fitting just one isochrone is insufficient, and a {\it multiple isochrone fitting} approach becomes necessary.

The underlying rationale is that a complex stellar population can be divided into sub-populations, each with a given age and chemical composition. Because these sub-populations with distinct ages and metallicities occupy different regions in the CAMD, one can simultaneously fit a set of isochrones, assigning each sub-population to a single isochrone. The set of isochrones is chosen {\it a priori} and may span a broad grid of ages and metallicities.

A maximum-likelihood scheme for {\it multiple isochrone fitting} was developed by \citet{small13}, who demonstrated that the CAMD of any observed stellar population can be ``{\it modeled as a linear combination of stellar isochrones}.'' The coefficients $\{a_{i}\}$ of this linear combination must satisfy $a_{i}\geq 0$ and $\sum_{i} a_{i}=1$, indicating that each $a_{i}$ can be interpreted as a probability weight. Consequently, larger $a_{i}$ values imply that the corresponding isochrone plays a more significant role in explaining the observed CAMD.

In the following subsections, we present a new methodology that builds on this principle but uses Bayesian inference instead of simply maximizing the likelihood function. This approach is more flexible and allows, among other things, to introduce physically motivated priors to constrain the solution when data alone cannot. 

\subsection{Bayesian inference}\label{sec:BayesInfer}

\citet{alzate21} employed Bayes’s theorem to compute the posterior probability distribution function (PDF) of $a_{i}$. We provide a concise overview of the method here.

At the core of this inference procedure is a predictive stellar population model that is integrated into Bayes's theorem.
The model requires the following fixed inputs:
\begin{enumerate}
    \item A set of isochrones predicting stellar brightness in a given photometric system.
    \item A density profile specifying how stars are distributed in heliocentric distance.
    \item An initial mass function (IMF), normalized by $\int \phi(M)~dM = 1$, which governs the distribution of stellar masses.
\end{enumerate}
We aim to sample the posterior PDF $p(\bm{a}, \bm{\beta} ~\vert~ \bm{d})$, where $\bm{\beta}$ represents the so-called nuisance parameters of the model (e.g., the theoretical distances $\rho$ and photometry $J^{k}$ of the stars), and $\bm{d}$ denotes the \textit{Gaia}/J-PLUS observational data.
Table~\ref{tab:statmod} describes the data and parameter sets used in our model.
Note that, rather than fitting the CAMD directly, our statistical model fits both the CMD and stellar parallaxes. For this reason, a distance prior is also required, not just the isochrones.
Because our principal goal is to infer the star formation history through the weights ${\bm a}$, we focus on the marginal posterior distribution:

\begin{eqnarray}\label{eq:bayes}
    p(\bm{a}~\vert~\bm{d}) \;=\; p(\bm{a}) \prod_{n=1}^{N_{D}} \int \frac{S(d_{n})~p\bigl(d_{n} ~\vert~ \beta_{n}\bigr)~p\bigl(\beta_{n}~\vert~ \bm{a}\bigr)}{\ell(S,\bm{a})}\; d\beta_{n},
\end{eqnarray}

\noindent
where $n$ indexes each star in the sample, $p(\bm{a})$ is the prior PDF on the weights, $p\bigl(\beta_{n} ~\vert~ \bm{a}\bigr)$ is the prior PDF on $\beta_{n}$ given $\bm{a}$, and $p\bigl(d_{n} ~\vert~ \beta_{n}\bigr)$ is the likelihood of the data for the $n^{\mathrm{th}}$ star. Appendix~\ref{app:stat_equ} details the forms of these prior PDFs and the likelihood function.

\begin{table}[ht]
\renewcommand{\arraystretch}{1.25} 
\begin{center}
 \caption{\label{tab:statmod}Variables entering our hierarchical model. Subscripts $i=1,...,I$ and $n=1,...,N$ indicate the $i^{th}$ isochrone  and the $n^{th}$ star, respectively, where $I$ is the number of selected isochrones and $N$ the number of observed stars. Superscript $k=1,2,...,K$ refers to the J-PLUS photometric filters as indicates in Table~\ref{tab:filters}.}
\resizebox{9cm}{!}{
\begin{tabular}{llcl}
 \hline
 Pop. parameter                    &             $\bm{a}$              &        &    Stellar fraction vector           \\
 \hline
 \multirow{4}{*}{parameters ($\bm{\beta}$)}       &              $\rho_{n}$              &   pc   &    Theoretical heliocentric distance \\
                                   &            $J_{n}^{k}$            &   mag  &    Theoretical Absolute magnitude    \\
                                   &            $\hat{\varpi}_{n}=1/\rho_{n}$     &   mas  &    Predicted parallax                \\
                                   &          $\hat{m}_{J,n}^{k}=J_{n}^{k}+f(\rho_n)$      &   mag  &    Predicted apparent magnitude      \\
 \hline
 \multirow{4}{*}{Data ($\bm{d}$)}             &            $\varpi_{n}$           &   mas  &    Observed parallax                 \\
                                   &           $e_{\varpi,n}$          &   mas  &    Parallax error                    \\
                                   &           $m_{J,n}^{k}$           &   mag  &    Observed apparent magnitude       \\
                                   &           $e_{J,n}^{k}$           &   mag  &    Apparent magnitude error          \\
 \hline
 \multirow{4}{*}{Fixed quantities} &             $p(\rho)$            &   mas  &    Distance prior                    \\
                                   &        $\mathcal{J}_{i}^{k}$      &   mag  &    Isochrone absolute magnitude      \\
                                   &          $\sigma_{i}^{k}$         &   mag  &    Isochrone tolerance               \\
                                   &              $\phi(M)$            &        &    Initial mass function             \\
 \hline
 \end{tabular}
}
\end{center}
\end{table}

The factor $\ell(S,\bm{a})$ in Eq.~(\ref{eq:bayes}) ensures proper normalization of the integral when accounting for the selection function $S$. It is calculated by

\begin{equation}\label{eq:selection}
    \ell(S,\bm{a}) \;=\; \int S\bigl(\bm{d}'\bigr)~ p\bigl(\bm{a}, \bm{\beta} ~\vert~ \bm{d}'\bigr)\; d\bm{\beta}~ d\bm{d}',
\end{equation}

\noindent
where $\bm{d}'$ is an auxiliary theoretical variable introduced to compute this normalization constant. 

Because our sample is 90\% complete in the magnitude range $m_{\rsds} = [14, 16.5]$~mag, we simplify by setting $S = 1$ for all $d_{n}$ in \jpxgf\ and restricting the integration limits in Eq.~(\ref{eq:bayes}) to $14 \leq J^{3}+\dm(\rho) \leq 16.5$. Failing to properly account for this completeness limit leads to incorrect predictions of stellar counts and consequently biases in the inferred SFH.

With all components of Bayes’ Theorem defined and restricted by the completeness limits, we sample the marginal posterior distribution $p(\bm{a}~\vert~\bm{d})$ in Eq.~(\ref{eq:bayes}) using a Monte Carlo method. The mode of $a_i$ provides the best estimate of the contribution of the $i$-th isochrone, while the 10\% and 90\% percentiles delineate the corresponding credibility interval.


\subsection{Stellar evolution model}\label{sec:stevol}

\begin{figure}
    \centering
    \includegraphics[width=1.\linewidth]{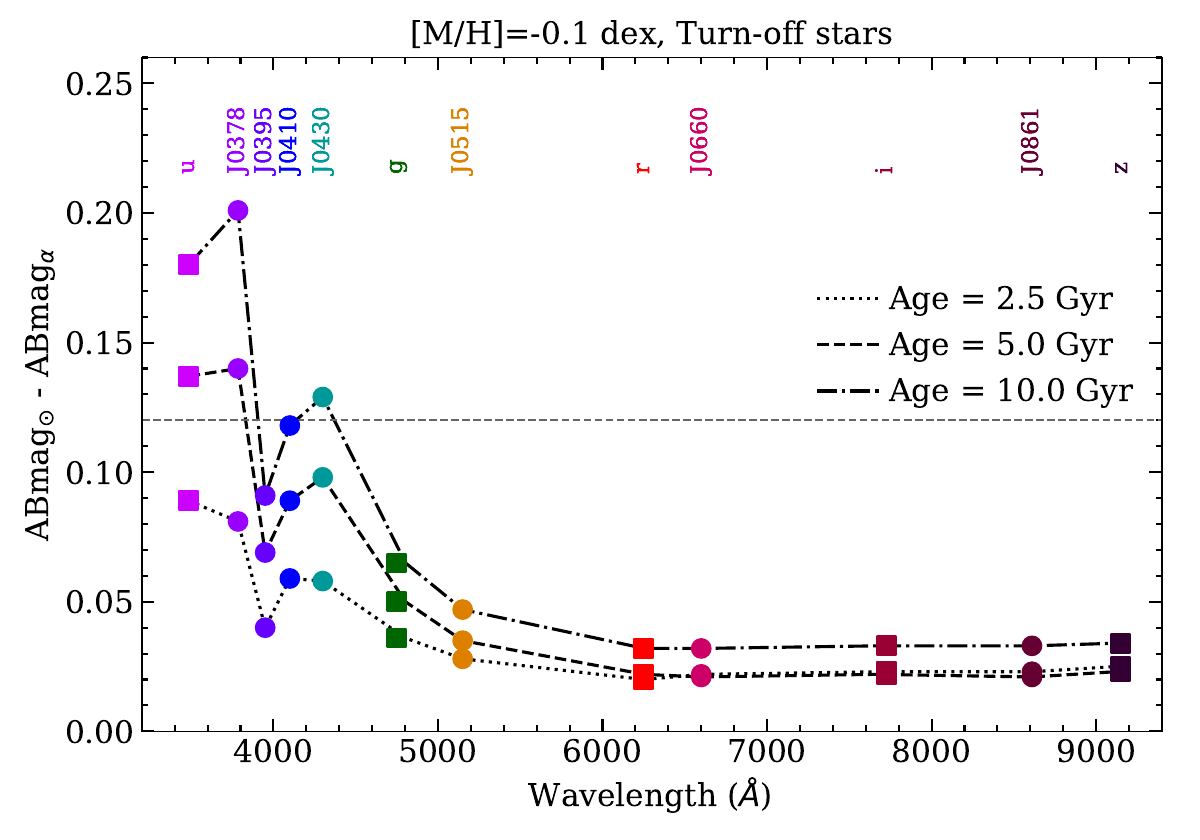}
    \caption{J-PLUS magnitude differences between solar-scaled and $\alpha$-enhanced BaSTI models for MSTO stars.
    We selected three stellar pairs at $\mh = -0.1$ dex, with approximate ages of 2.5, 5.0 and 10 Gyr, and corresponding masses of 1.3, 1.1, and 1.0 M$_\odot$, respectively.
    The horizontal dashed line indicates the upper limit of the median absolute magnitude error, as averaged from the third column of Table~\ref{tab:stats}.
    }
    \label{fig:sol_alpha}
\end{figure}

\begin{figure}
    \centering
    \includegraphics[width=0.9\linewidth]{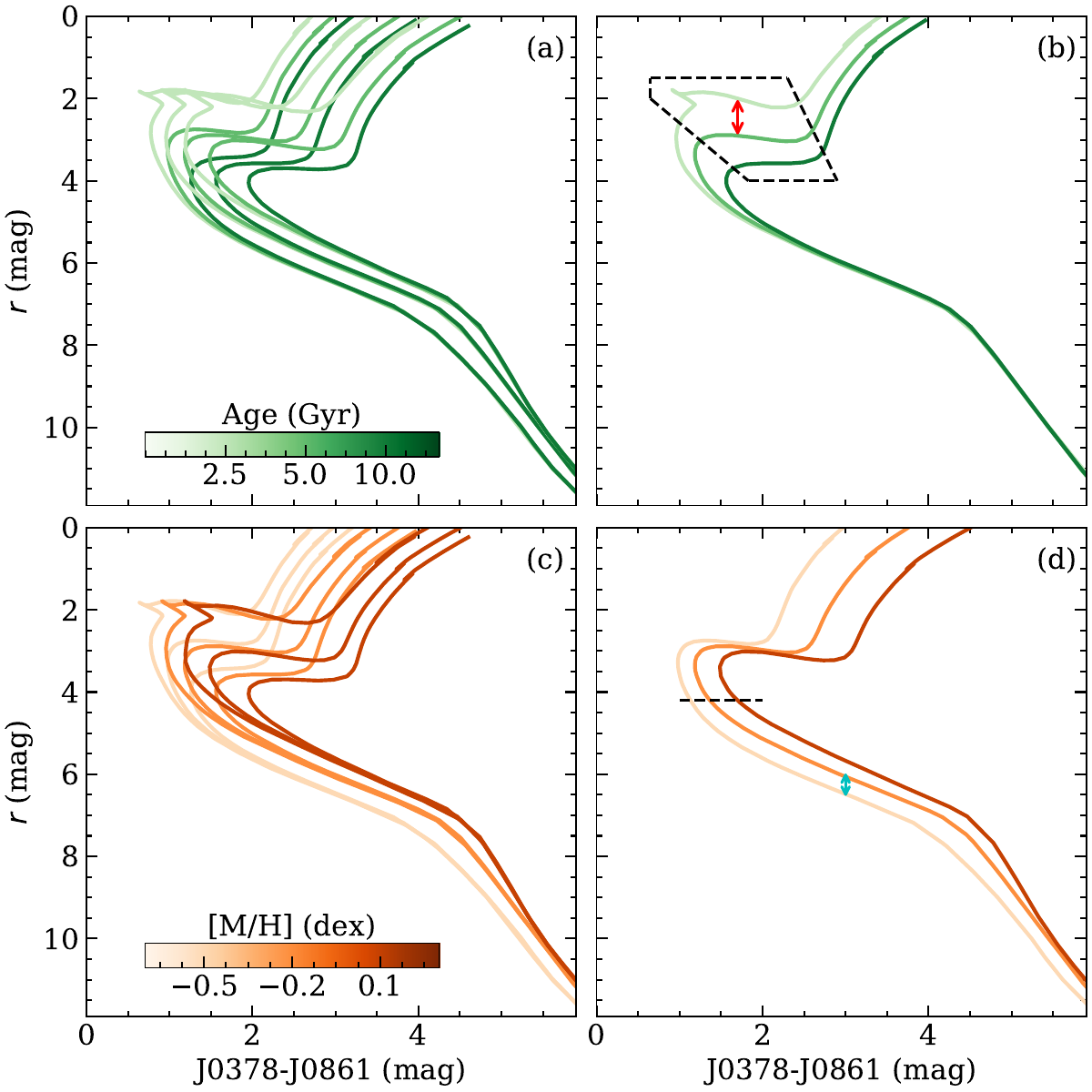}
    \caption{Illustrative isochrone plot. Panel (a) displays a grid of 9 isochrones with ages of $2.5$, $5$, and $10$ Gyr, and metallicities of $\mh = -0.5$, $-0.2$, and $0.1$ dex.
    Isochrones with the same age are shown in green tones, illustrating the sensitivity of age and metallicity at the MSTO and the RGB phases.
    Panel (b) highlights the region of the CAMD where age sensitivity is highest ( dashed polygon), using three isochrones of the same metallicity ($\mh=-0.2$ dex).
    The double-headed red arrow indicates the magnitude separation used to quantify the average distance between isochrones.
    Panel (c) shows the same grid as (a), but with orange tones indicating isochrones of the same metallicity.
    Panel (d) shows three isochrones of the same age (5 Gyr), illustrating how magnitude and colour vary with metallicity. The cyan arrow indicates the magnitude separation used to define the average distance between isochrones.
    }
    \label{fig:cmdgrid}
\end{figure}

\begin{table}
\caption{\label{tab:iso} Isochrone Grids used in this work. The step in metallicity is $\Delta$[M/H]$= 0.1$ dex for all the grids.}
\centering
 \resizebox{9cm}{!}{
 \setlength{\tabcolsep}{2pt}
\begin{tabular}{ccccccccc}
\hline
\multirow{2}{*}{Grid}  & \multirow{2}{*}{Evol. model}    & $\log$(age/yr)    & $\Delta$ $\log$(age)     &    [M/H]     &  $[\alpha/{\rm Fe}]$  & \multirow{2}{*}{N$_{\rm iso}$} \\
     &               &        (Gyr)     &      (Gyr)       &     (dex)      &   (dex)   &         \\
\hline
 A   &  PARSEC   &  9.3, 9.7, 10.0  &  $\approx 0.35$  &  $[-1.5,0.3]$  &   $0.0$   &   57    \\
 B   &  PARSEC   &   $[8.7,10.1]$   &      $0.05$      &  $[-1.5,0.3]$  &   $0.0$   &   551   \\
\hline
 C   &  BaSTI    &   $[8.7,10.1]$   &      $0.05$      &  $[-1.5,0.3]$  &   $0.0$   &   551   \\
 D   &  BaSTI    &   $[8.7,10.1]$   &      $0.05$      &  $[-1.5,0.3]$  &   $0.4$   &   551   \\
 E   &  BaSTI    &   $[8.7,10.1]$   &      $0.05$      &  $[-1.5,0.3]$    &$[0.0,0.4]$&   1102  \\
\hline
\end{tabular}
}
\end{table}

We employ two stellar evolution libraries that incorporate the J-PLUS photometric system.
First, the BaSTI-IAC stellar evolution models provide updated sets of solar-scaled \citep{hidalgo18} and $\alpha$-enhanced \citep{pietrin21} evolutionary tracks, covering \feh from $-3.20$ to $+0.45$ dex and $-3.20$ to $+0.06$ dex, respectively, and spanning stellar masses from $0.1$ to $15$ M$_{\odot}$.
The mass step depends on the stellar mass, ranging from $\Delta M = 0.05~\msun$ to $1~\msun$, while the metallicity step varies from $\Delta$\feh$ = 0.08$~dex to $0.7$~dex for solar-scaled models, and from $\Delta$\feh$ = 0.15$–$0.20$~dex for $\alpha$-enhanced models.
Both grids include isochrones for ages ranging from $20$ Myr up to $14.5$ Gyr.
The solar-scaled models adopt the elemental metal distribution from Caffau et al. (2011). For the alpha-enhanced models, while the non-$\alpha$ elements are maintained as Caffau's mixture, the $\alpha$-elements O, Ne, Mg, Si, S, Ca, and Ti are uniformly enhanced with respect to Fe by a fixed ratio of \afe=0.4, increasing the the total metallicity $\mh$ at a given \feh.
Although their evolutionary tracks follow similar trends, notable differences appear in their photometry.
In particular, isochrones with $\alpha$-enhancement tend to be slightly bluer in the B–V colour by about $0.04$ mag compared to their solar-scaled counterparts,
especially in the blue and ultraviolet passbands, whereas differences in the infrared and $V-I$ colour–magnitude diagrams are less significant.
This effect can also be observed in the J-PLUS photometric system, where filters bluer than 5000 {\AA} show the largest differences in flux,
making them highly sensitive to $\alpha$-abundance variations (see Fig.~\ref{fig:sol_alpha}).
The BaSTI-IAC models account for diffusive processes, core convective overshooting, and mass loss (Reimers’ parameter $\eta_{\rm Reimers} = 0.3$).

The second library is PARSEC1.2S \citep{bressan12, tang14, chen15}, accessed through its public web interface\footnote{\url{http://stev.oapd.inaf.it/cgi-bin/cmd}}.
This version assumes solar chemical mixture and provides stellar evolutionary tracks covering a wide metallicity range ($-2.2 \leq \mh \leq +0.5$) and initial masses from 0.1 up to 150 $M_\odot$, with extensions reaching 350 $M_\odot$ for subsolar metallicities (\mh~$\leq 0.15$ dex). The models include key physical processes such as microscopic diffusion, convective core overshooting, and mass loss following the Reimers formulation with an efficiency parameter $\eta = 0.3$. Isochrones derived from these tracks are computed from the pre-MS up to advanced evolutionary phases.

\subsection{Sampling the CAMD}\label{ssec:samp_cmd}

Stars in a CAMD are arranged according to intrinsic properties such as effective temperature, surface gravity, age, and chemical composition. Certain regions in the CAMD are particularly sensitive to these parameters, whereas others are not. This requires a well-sampled CAMD to maximise the information extracted from evolutionary phases that are most sensitive to the star formation history.

Panels (a) and (c) in Fig. 4 illustrate how isochrones of different ages and metallicities are distributed in the CAMD.
Regions covering the MSTO and the more evolved phases of stellar evolution are those containing the most information about the age of the stars, but they are subject to a strong age-metallicity degeneracy. In contrast, the low-mass MS phase is notably insensitive to age but not to metallicity. To exploit this difference, we split the CAMD at $\rsds = 4.2$~mag: the upper region ($\rsds \leq 4.2$~mag) is used to derive the SFH, while the lower region ($\rsds > 4.2$~mag) supplies a metallicity prior that helps mitigate possible age--metallicity degeneracies in the upper portion of the diagram.

Before assembling our isochrones grids, we assessed the minimum metallicity and age that can be reliably distinguished given our magnitude uncertainties (see Table~\ref{tab:stats}).
Specifically, we defined an isochrone separation $\Delta J^k$ as the average difference in absolute $J^k$ magnitude between two isochrones sharing the same age (or metallicity) but adjacent metallicities (or ages).
This approach is illustrated in Fig.~\ref{fig:cmdgrid}: for example, the double-headed arrow in panel (b) marks the $\rsds$ separation at $(J0378 - J0861) = 1.7$~mag around the MSTO and SGB phases.
In panel (d), we measure a similar separation  at $(J0378 - J0861) = 3.0$~mag, computed for $\rsds \geq 4.2$~mag and stellar masses $>0.2~\msun$.
Our isochrone sets must satisfy $\Delta J^{k} >$ the average absolute magnitude errors ($\approx 0.12$~mag), ensuring that any differences recovered in the posterior distribution of $a_i$ reflect actual variations in the CAMD rather than noise.

Our final set of models comprises grids from both BaSTI-IAC (for $\alpha=0$ and $\alpha=0.4$) and PARSEC, spanning ages from $\sim$13.5~Gyr to 500 Myr in logarithmic steps, and metallicities in the range $-1.5 \leq \mh \leq 0.3$. For consistency with our observations, we exclude Helium-burning (HB) and asymptotic giant branch (AGB) because they are out of our completeness interval, and we limit our analysis to MS, SGB, and red giant branch (RGB) stars. Additionally, we remove pre-MS stars and adopt $\eta_{\rm Reimers} = 0.3$ to quantify mass loss on the RGB.
Table~\ref{tab:iso} summarizes all the grids used in this work.

Another important factor is the isochrone tolerance, $\sigma$, which defines the width of the probability distribution for each isochrone’s location in the CAMD. Although a non zero $\sigma$ can improve numerical stability in cases of very small photometric errors, our absolute magnitude uncertainties (dominated by distance modulus errors) are $\approx 0.12$~mag for all J-PLUS filters, making the effect of $\sigma$ negligible (\(\sigma^{k}_{i} \ll \Delta J^{k}\)). Hence, the choice of $\sigma$ does not significantly influence the final SFH solutions in this study.

\subsection{Building the SFH}

We corrected the bias in the inferred $a_i$ values caused by the absolute magnitude cut-off at $\rsds = 4.2$ mag. For $\rsds \leq 4.2$ mag, the resulting $a_i$ values are underestimated due to the absence of fainter stars. In this case, the correction factor is
\begin{equation}
    f_{C}=\left[\int_{0.1~\msun}^{M(\rsds=4.2) } \phi(M) dM\right]^{-1},
\end{equation}
\noindent and, when $\rsds>4.2$ mag, the correction is 
\begin{equation}
    f_{C}=\left[\int_{M(\rsds=4.2)}^{100~\msun} \phi(M) dM\right]^{-1}.
\end{equation}

After the correction, we re-normalize the $a_i$ and derive the SFH $\psi(\mh,{\rm age})$ as follows
\begin{equation}
    \psi(\mh_{i},{\rm age}_{i})=\frac{a_i}{\Delta\mh\cdot\Delta{\rm Age}_{i}},
\end{equation}
\noindent where $\Delta\mh=0.1$ dex, $\Delta{\rm age}_{i}={\rm age}_i\cdot(10^{\Delta_{\rm \log{\rm age}}}-1)$ Gyr and $\Delta\log({\rm age})=0.1$ dex. The marginal distributions are given by $\psi(\mh_i) = \sum_j \psi\bigl({\rm age}_j,~\mh_i \bigr)\cdot \Delta{\rm age}$ and $\psi({\rm age}_i) = \sum_j \psi\bigl({\rm age}_i,~\mh_j \bigr)\cdot \Delta\mh$. The distributions are normalized such that $\sum_i \psi\bigl(\mathrm{Age}_i,~[\mathrm{M/H}]_i \bigr)\cdot\Delta\mh\cdot\Delta{\rm Age}_{i} = 1$.

\subsection{Mitigating degeneracies with multi-filter photometry}

J-PLUS photometry has demonstrated significant potential in mitigating the age–metallicity degeneracy commonly encountered in CAMD fitting.
The use of multiple narrow and broad filters, strategically designed to target crucial stellar absorption features, enhances its sensitivity to elemental abundances \citep{cenarro19}.
Specifically, filters such as $u$, $z$, $J0378$, and $J0861$ broaden the spectral baseline, thereby diminishing degeneracies between colour (or effective temperature) and metallicity, resulting in more precise determinations of {\feh}.
Additionally, the $J0395$ filter, centred on the Ca II H and K lines, strongly correlates with iron abundance, increasing the robustness of metallicity measurements.
As described in Section 3, stars fainter than the MSTO ($\rsds > 4.2$ mag) are particularly effective in tracing the metallicity distribution within the \jpxgf sample, enabling the construction of an age-independent metallicity prior.
This prior significantly reduces degeneracies for brighter stars ($\rsds \leq 4.2$ mag).
Grid A in Table~\ref{tab:iso} was defined to fully exploit these properties.

The chemical sensitivity provided by the J-PLUS photometric system, in conjunction with $\alpha$-enhanced isochrones, allows for consistent CAMD fitting of stellar populations with enhanced $\alpha$-abundances.
Filters such as $J0515$, centred on the Mg b triplet, and $J0861$, which covers the Ca II triplet \citep{cenarro19}, serve as indicators of magnesium and calcium relative to iron, making them useful for constraining $\alpha$-element abundances.
Additional filters sampling the G band ($J0430$) and the hydrogen Balmer lines ($u$) further enable J-PLUS to accurately constrain stellar atmospheric parameters and key abundance ratios, including [Mg/Fe] and \afe, as evidenced by prior studies \citep{whitten19,yang22,huang24}.
The $\alpha$-enhanced BaSTI models assume a uniform enhancement pattern of \afe = 0.4 dex across elements O, Ne, Mg, Si, S, Ca, and Ti \citep{pietrin21}.
Notably, the J-PLUS blue filters ($<5000$\AA) exhibit sensitivity to variations in $\alpha$-enhancement levels within isochrones (Sec.~\ref{sec:stevol}), enabling the effective separation of high-$\alpha$ and low-$\alpha$ populations in the CAMD.
The agreement between the sensitivity of J-PLUS photometry and theoretical stellar evolution models to different $\alpha$-element compositions, validates the application of isochrone grids for studying $\alpha$-enhanced populations, such as those found in the Galactic thick disc.

\subsection{Estimating uncertainties}
Grids B, C, and D span a broader range of ages and metallicities, increasing the risk of obtaining biased SFHs due to age–metallicity degeneracies. To evaluate the impact of random uncertainties and degeneracies on the age and metallicity distributions, we performed recovery tests over a synthetic grid of simple stellar populations (SSPs). This synthetic population include realistic photometric and parallax uncertainties characteristic of the JxG5 sample (see Appendix~\ref{app:mock_test}).

These tests showed that the uncertainty distribution depends significantly on the SSP age and metallicity,
clearly revealing the effect of parameter degeneracies.
By fitting Normal and Cauchy distribution functions to the marginal distributions of $\mh$ and age,
we estimate typical uncertainties of $\sigma_{\mh}\approx 0.10 - 0.14$ dex,
and $\sigma_{\rm age}\approx0.15$ Gyr for young/intermediate ages and $\sigma_{\rm age}\approx0.55 - 1.2$ Gyr for old ages,
associated the \jpxgf sample.

These uncertainties are in good agreement with the typical values obtained by \citet{huang24}, who reported metallicity uncertainties of $0.10$–$0.20$ dex using kernel-based regression applied to J-PLUS/{\g} colours, and 20\% uncertainties for stellar ages determined from isochrone fitting of MSTO and SGB stars. We did not perform a similar uncertainty estimation for $\alpha$-abundance determinations, given our model limitation to discrete values of \afe=0.0 and 0.4 dex.


\section{Results}\label{sec:results}

\subsection{Metallicity distribution of the MS low-mass stars}

\begin{figure}
   \centering
   \includegraphics[width=0.39\textwidth]{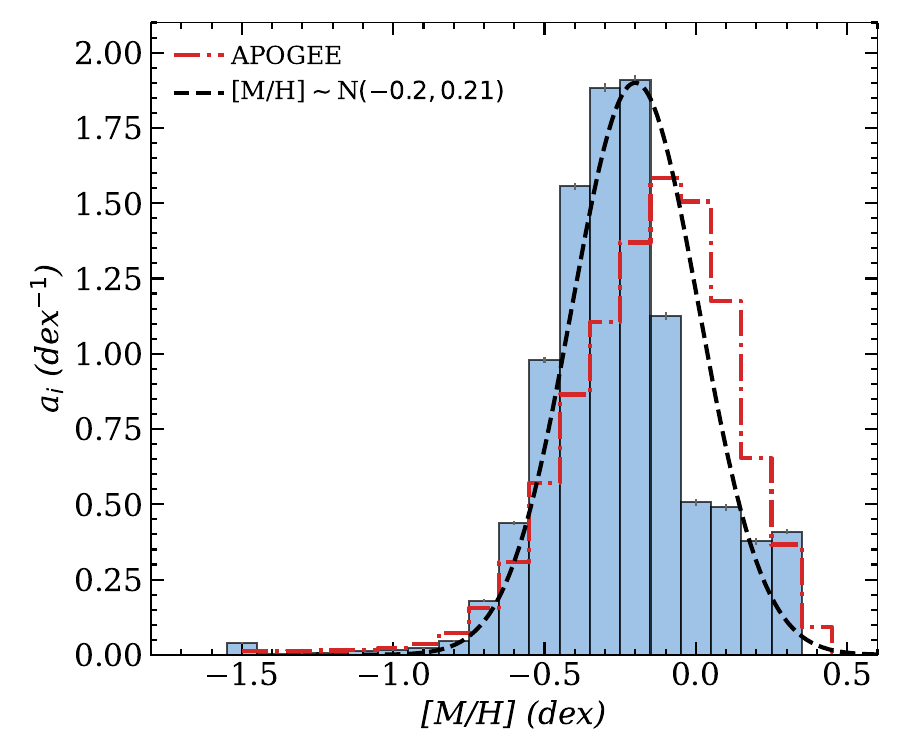}
   \caption{
   Inferred metallicity distribution of the \jpxgf\ sample (solid histogram) for stars with absolute $\rsds$ magnitudes fainter than 4.2 mag, using Grid A from Table\ref{tab:iso}.
   The dash-dotted line shows the metallicity distribution of an APOGEE DR17 sample, selected as described in Appendix~\ref{app:apo_samp}.
   The dashed line indicates the normal function adopted as the prior on $\mh$ for the PARSEC models.
   }
   \label{fig:mh_jplus}
\end{figure}

Using Grid A in Table \ref{tab:iso}, we derived the metallicity distribution for the \jpxgf stars that are fainter than $\rsds = 4.2$ mag. Grid A contains 57 isochrones organised into three age bins ($\sim2$, 5, and 10~Gyr), each bin containing 19 metallicity steps. We limited the ages to these three values because the faint part of the CAMD ($\rsds > 4.2$ mag) is insensitive to age changes (see the top-left panel of Fig.~\ref{fig:cmdgrid}).
Fig.~\ref{fig:mh_jplus} shows the resulting metallicity distribution. It peaks at $[M/H] \sim -0.2$ dex, and 90\% of the stars lie between -1.0 and 0.4 dex.

In contrast to the metallicity distributions obtained from APOGEE DR17, the result reveals that fits based on the PARSEC isochrones produce a reasonable distribution but tend to recover metallicities that are $\sim0.1$ dex lower than spectroscopic values.
Possible explanations for this 0.1 mag shift include, underestimated extinction values from \texttt{BAYESTAR} and/or the possibility that the PARSEC models predict slightly cooler effective temperatures for low-mass main sequence stars.
The 1\% accuracy of the J-PLUS photometric calibration is unlikely to account for a shift of 0.1 mag \citep{sanjuan24}.

Compared to the BaSTI solar-scaled models, the PARSEC~1.2S models proved more effective at fitting the low-MS stars photometry of the \jpxgf. This may be attributed to the temperature–optical depth ($T$–$\tau$) relations from PHOENIX BT-Settl atmospheres, as implemented by \citet{chen14}. In contrast, the BaSTI models for low-mass stars are significantly bluer than the J-PLUS colours and did not yield a metallicity distribution with well-defined features. For this reason, they were not used to define the prior.
A similar offset was found for the PARSEC set by \citet{gallart24} using \g~data, but not for the BaSTI tracks.

On the other hand, from test analyses conducted on the bright part ($\rsds\leq4.2$ mag) of the \jpxgf sample, we found that the obtained metallicity distribution is systematically biased towards higher metallicities compared to those reported by spectroscopic surveys. This effect can be observed in \citet{gallart24} as well, and can be produced by the inherent age-metallicity degeneracy of the isochrones, the presence of unresolved binary stars or even limitations in the stellar evolutionary models.

Because the metallicity distribution in the lower CAMD is essentially age–independent, we use it to
define informative priors that help to mitigate the age–metallicity degeneracy when modelling the
upper CAMD:
\begin{align}
  \text{PARSEC:}\quad & [{\rm M/H}] \sim \mathcal{N}(-0.2,\,0.21)\;\text{dex}, \label{eq:prior_parsec}\\
  \text{BaSTI:}\quad & [{\rm M/H}] \sim \mathcal{N}(-0.1,\,0.21)\;\text{dex}, \label{eq:prior_basti}
\end{align}

\noindent where the BaSTI prior adopts the same dispersion but is shifted by $+0.10$ dex, following
the offset found in Fig.~\ref{fig:mh_jplus} for the metallicity distributions between the spectroscopic data and our result.
Imposing these priors during the inference process suppresses spurious high-metallicity
solutions and yields more reliable age determination.

\subsection{Star formation history: solar-scaled and $\alpha$-enhanced models}\label{ssec:sfh}

\begin{figure}
    \centering
    \includegraphics[width=0.95\columnwidth]{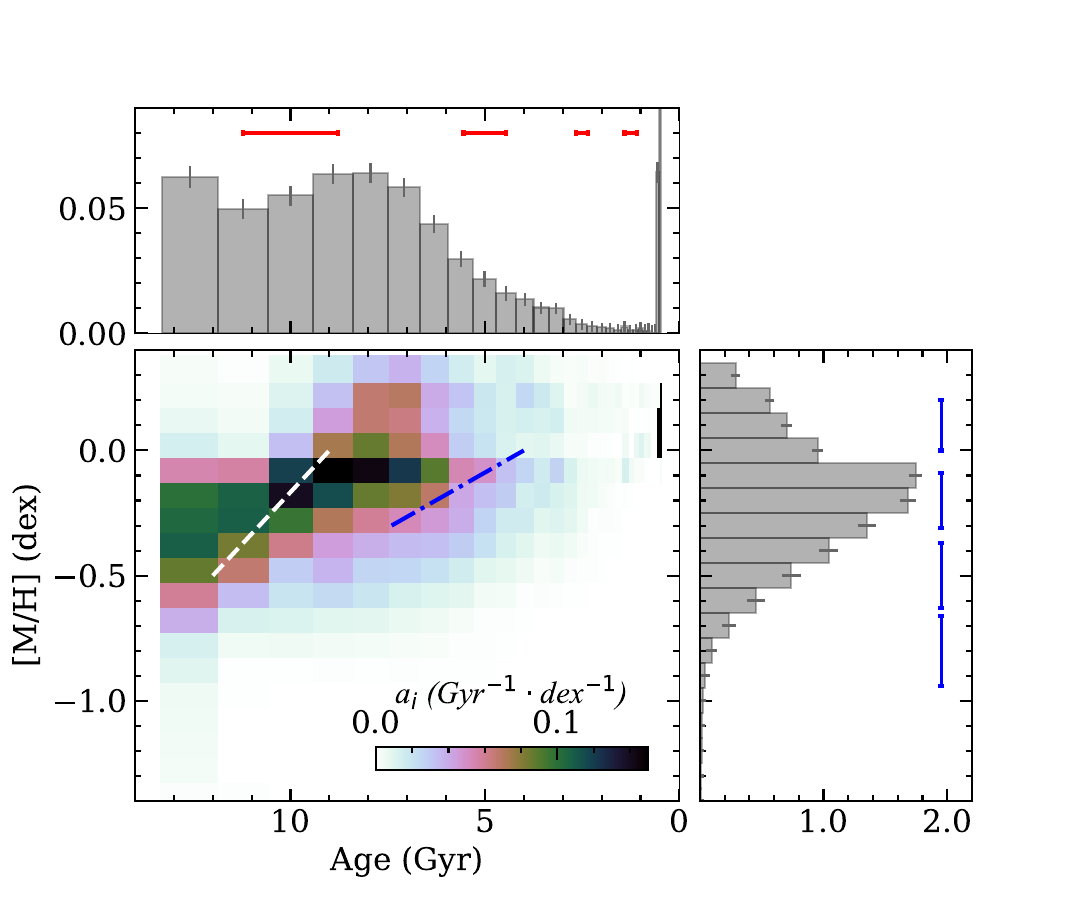}
    \caption{Inferred SFH of the \jpxgf sample for stars brighter than $\rsds=4.2$ mag, using Grid B. The grey histograms in the upper and right panels show the marginal distributions of age and metallicity, respectively. The red and blue error bars above the histograms indicate the full width at half maximum (FWHM) of the age and metallicity uncertainties, as estimated in Appendix~\ref{app:mock_test}. The dashed and dash-dotted lines highlight the two population candidates that we consider to be detected.}
    \label{fig:amr_parsec}
\end{figure}

\begin{figure*}
    \centering
    \includegraphics[width=0.9\columnwidth]{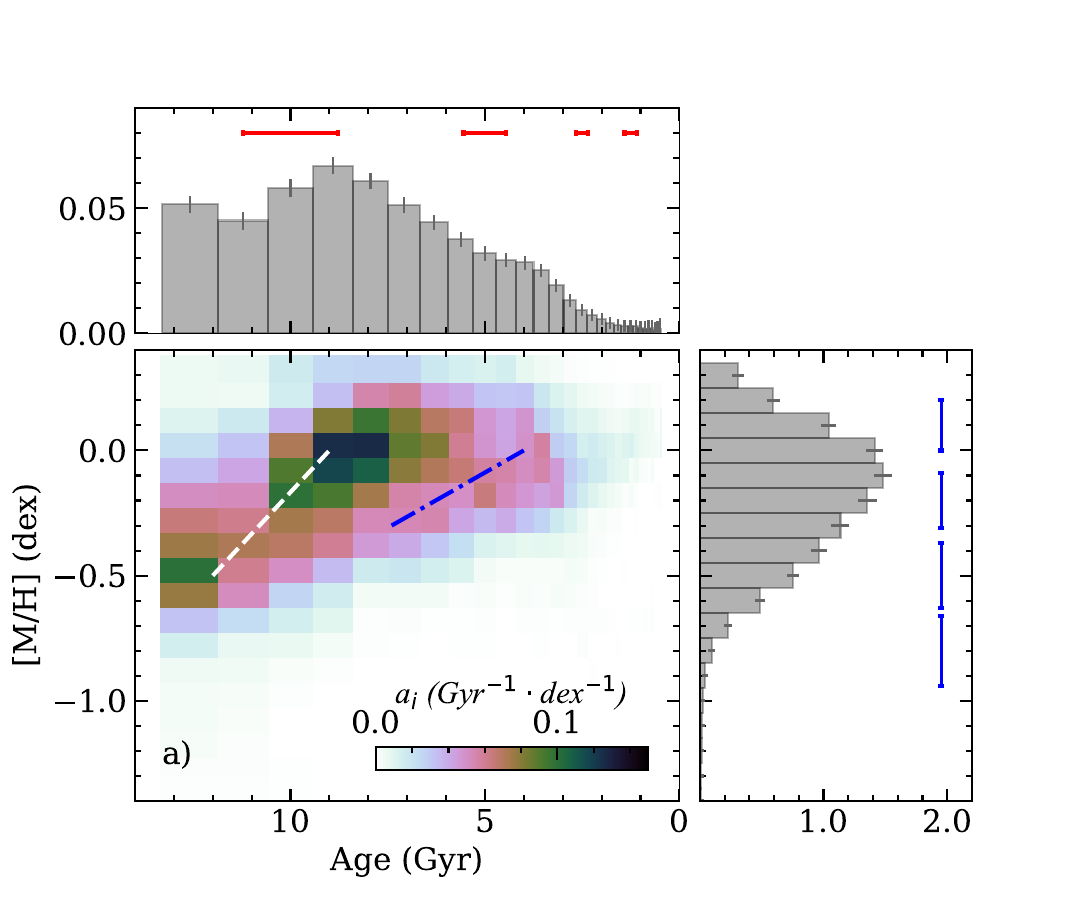}
    \includegraphics[width=0.9\columnwidth]{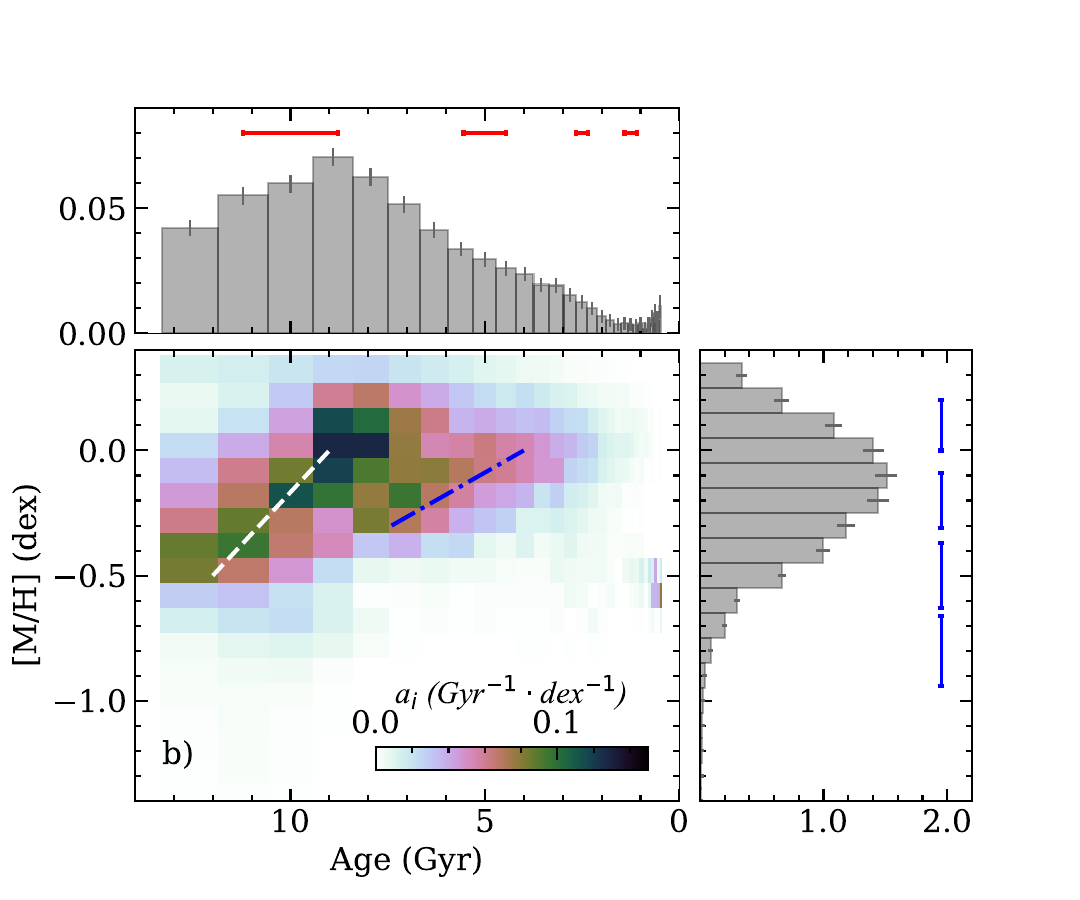}
    \caption{Inferred SFH of the \jpxgf sample for stars brighter than $\rsds=4.2$ mag, using Grid C (solar-scaled; panel a) and Grid D ($\alpha$-enhanced; panel b).
    The grey histograms in the upper and right panels show the marginal distributions of age and metallicity, respectively.
    The red and blue error bars above the histograms indicate the full width at half maximum (FWHM) of the age and metallicity uncertainties, as estimated in Appendix~\ref{app:mock_test}.
    The dashed and dash-dotted lines highlight the two population candidates that we consider to be detected.}
    \label{fig:amr_basti}
\end{figure*}

\begin{figure*}
    \centering
    \includegraphics[width=1.5\columnwidth]{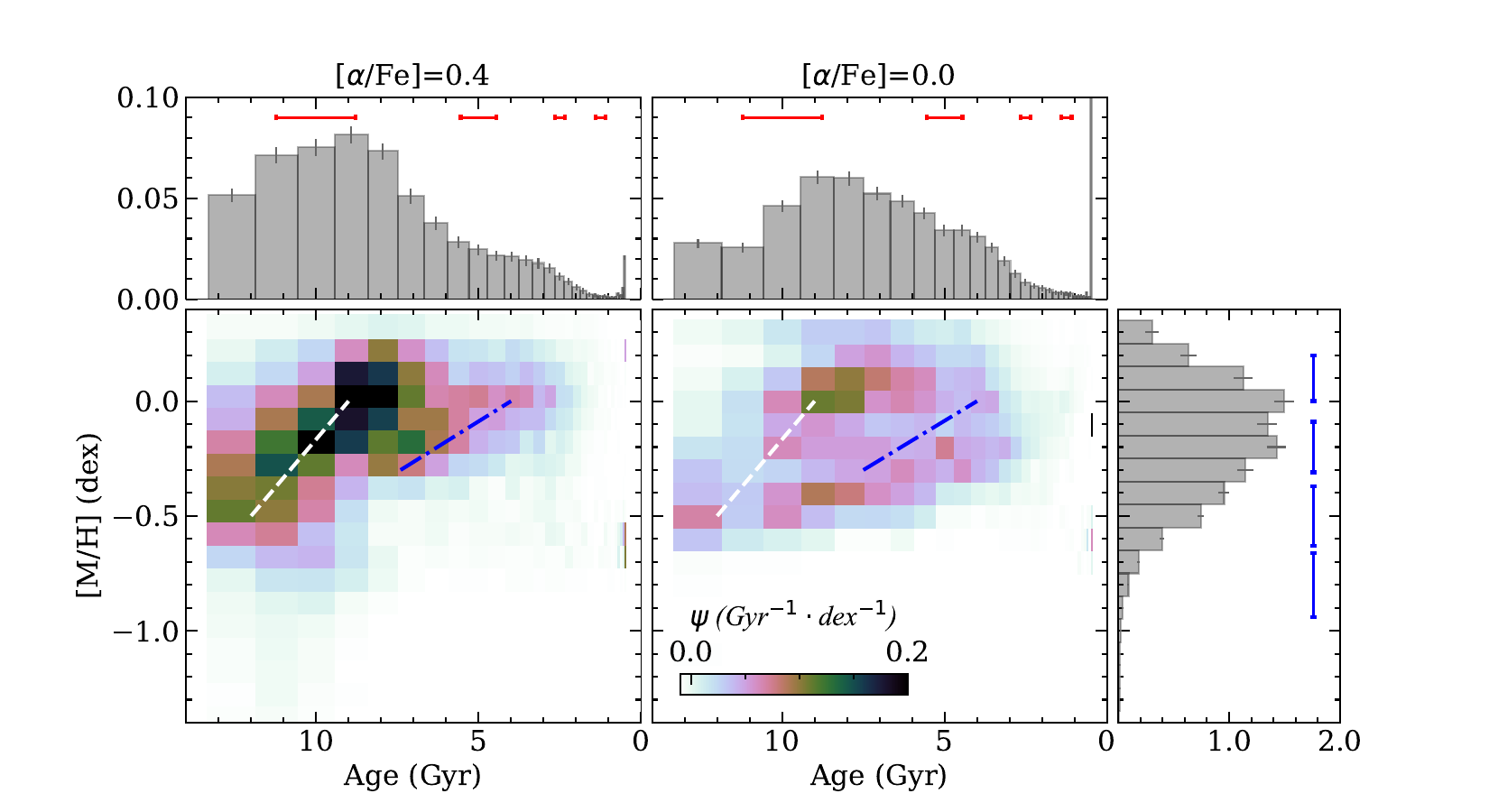}
    \caption{
    Inferred SFH of the \jpxgf sample for stars brighter than $\rsds=4.2$ mag, using Grid E, which simultaneously fits solar-scaled and $\alpha$-enhanced isochrones (indicated as \afe=0.0 and \afe=0.4, respectively, in the panel titles).
    The grey histograms in the upper and right panels show the marginal distributions of age and metallicity, respectively.
    The red and blue error bars above the histograms represent the full width at half maximum (FWHM) of the age and metallicity uncertainties, as estimated in Appendix~\ref{app:mock_test}. 
    The dashed and dash-dotted lines highlight the two candidate populations that we consider to be detected.
    }
    \label{fig:amr_basti2}
\end{figure*}

Figs.~\ref{fig:amr_parsec} and \ref{fig:amr_basti} show the inferred SFHs for the \jpxgf kpc sample ($M_r \leq 4.2$ mag), obtained using grids B, C, and D from Table~\ref{tab:iso}.
Grids B and C correspond to the SFHs inferred using PARSEC and BaSTI models with solar-scaled abundances, respectively, while Grid D represents the SFH inferred from BaSTI models with $\alpha$-enhanced abundances.
These two figures thus highlight the differences introduced by the adoption of different stellar evolution codes and $\alpha$-enhanced models.
The age axes in these figures are presented on a linear scale and the corresponding age uncertainty is approximately $\sigma_{\rm age}\approx0.15$ Gyr for young/intermediate ages and $\sigma_{\rm age}\approx0.55 - 1.2$ Gyr (Appendix \ref{app:mock_test}).

Figs.~\ref{fig:amr_parsec}, \ref{fig:amr_basti}a and \ref{fig:amr_basti}b 
reveal a stellar population exhibiting a monotonic chemical enrichment, with metallicity increasing from about $[M/H]\simeq -0.6$ to $-0.5$ dex at 12.5 Gyr up to approximately $0.0$ dex by 8 Gyr.
We highlight this population's chemical signature with a white dashed line,
corresponding visually to an enrichment rate of approximately $0.13~{\rm dex\cdot Gyr}^{-1}$.
This line serves solely as a visual guide rather than representing a formal fit.
Within our uncertainties, this chemical enrichment trend is consistent with age-metallicity relations derived by \citet{xiang22} from LAMOST, \citet{sahlholdt22} from GALAH, \citet{cerqui25} using APOGEE, and \citet{alvar25} using \g, specifically for the thick disc stellar population.
A second metallicity enrichment pattern is clearly visible in Fig.~\ref{fig:amr_basti}b, it begins at approximately $\mh=-0.3$ dex between 8 and 7 Gyr ago and reaches $\mh\simeq0.0$ dex by around 3 Gyr.
This second enrichment feature is also partially visible in Figs.~\ref{fig:amr_parsec} and \ref{fig:amr_basti}a, though with less clarity.
All our inferred SFH indicate the presence of stars older than 7 Gyr with metallicities $[M/H]\geq0.0$ dex.

\subsubsection{Simultaneous fitting of Solar-scaled and $\alpha$-enhanced models}\label{sssec:afe_solar}

Grid E from Table~\ref{tab:iso} includes both solar-scaled and $\alpha$-enhanced isochrones.
This grid allows our algorithm to determine which of the two $\alpha$-abundance regimes, or a combination of them, is more likely to fit the observed CAMD.
In Fig.~\ref{fig:amr_basti2}, we present the SFH derived from this grid, separating the distributions into the $\alpha$-enhanced (\afe = 0.4 dex) and solar-scaled (\afe = 0.0 dex) components.

The marginal age distribution in Fig.~\ref{fig:amr_basti2} shows that, for the $\alpha$-enhanced population, most stars formed between 12.5 and 8 Gyr ago. Conversely, marginal age distribution for the solar-scaled population, the contribution from older stars diminishes significantly, with intermediate-age populations dominating the distribution. Additionally, the SFH indicates that, for the $\alpha$-enhanced population, metallicity growth slowed down at around 9 Gyr, and an intermediate-age population emerges at about 8 Gyr.

The $\alpha$-enhanced SFH in Fig.~\ref{fig:amr_basti2} exhibits a chemical enrichment trend between 12.5 and 8 Gyr that closely matches those seen in Figs.\ref{fig:amr_parsec} and \ref{fig:amr_basti}, but with the $\alpha$-enhanced models dominating over the solar-scaled ones.
Metallicity growth then plateaus around 8 Gyr at $\mh \approx 0.1-0.2$ dex, after which a second enrichment episode commences in the $\alpha$-enhanced population.
The apparent overlap between these two enrichment phases can be attributed to the systematic age–metallicity correlation identified in our mock tests (Appendix~\ref{app:mock_test}).

In the solar-scaled SFH of Fig.~\ref{fig:amr_basti2}, there is a significant reduction in the number of stars older than $\sim$9–10 Gyr.
Moreover, between 10 and 3 Gyr, a secondary enrichment trend is apparent, with metallicity growing more slowly compared to that of the older, high-$\alpha$ population.
Another feature emerges at metallicities $\mh\geq0.0$ dex, spanning ages from about 8 Gyr down to approximately 3 Gyr. This second population appears to show a slightly decreasing trend in metallicity, evolving from about $(8~{\rm Gyr},0.2~{\rm dex})$ down to $(3~{\rm Gyr},0.0~{\rm dex})$. Lastly, a third population is visible at sub-solar metallicities ($\mh<0.0$ dex), showing a moderate increase in metallicity from approximately $\mh=-0.4$ dex between 9 and 8 Gyr ago, to about $\mh=-0.2$ dex at around 3 Gyr.

\subsection{Spatially mapping metallicity and age gradients}\label{ssec:zcorr}

\begin{figure}
    \centering
    \includegraphics[width=1.0\linewidth]{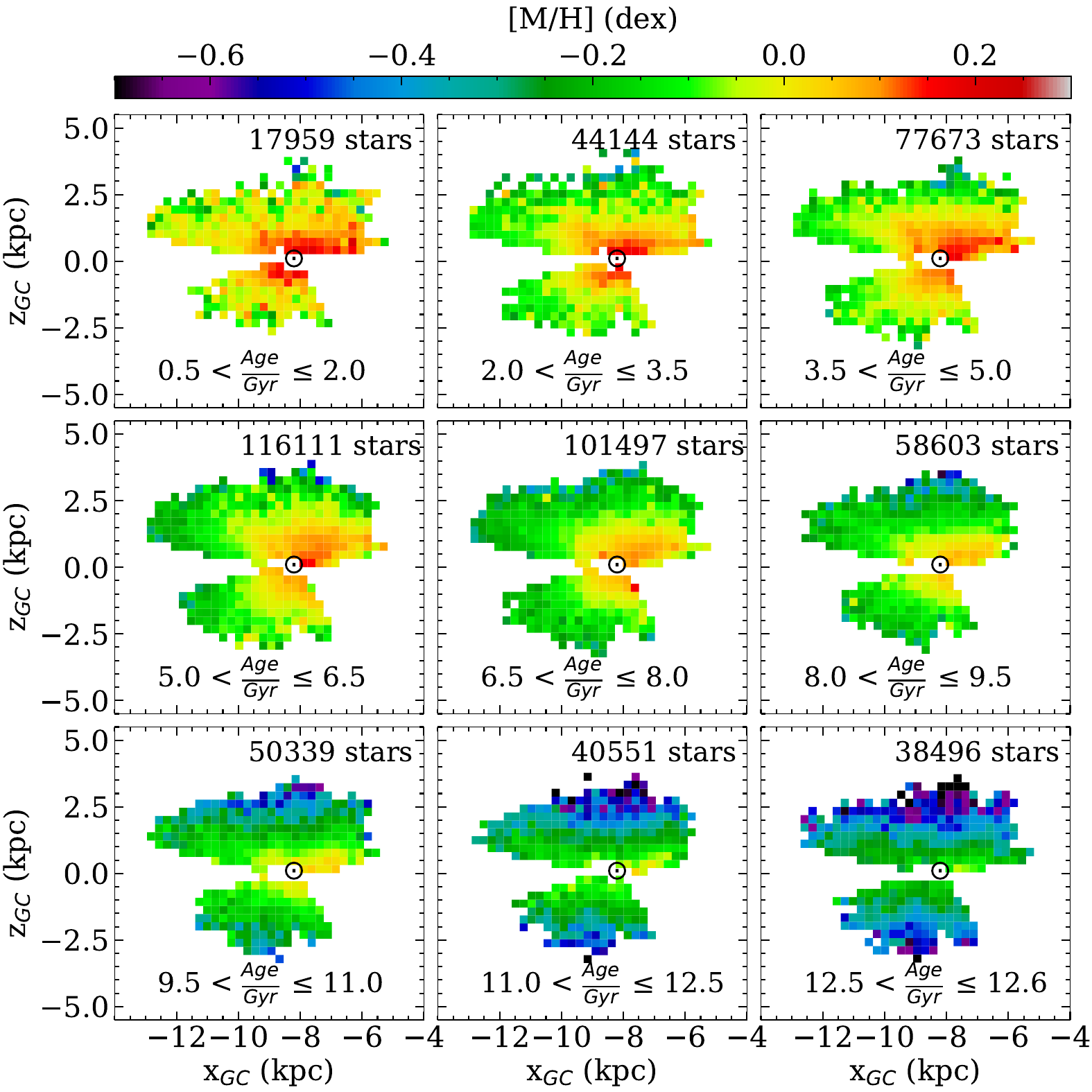}
    \caption{Metallicity map in galactocentric $x_{GC}$ and $z_{GC}$ coordinates of the \jpxgf ($\rsds\leq 4.2$ mag) sample.
    The metallicity values were derived as described in Sec.~\ref{ssec:zcorr}, using Grid E (solar+$\alpha$-enhanced isochrones) from Table~\ref{tab:iso}.
    The colour-bar indicates the mean of $\mh$ for stars within $0.2\times0.2$ kpc$^2$ pixels.
    The pixels with less than 10 stars were rejected.
    }
    \label{fig:xz_mean_mh}
\end{figure}

\begin{figure}
    \centering
    \includegraphics[width=1.0\linewidth]{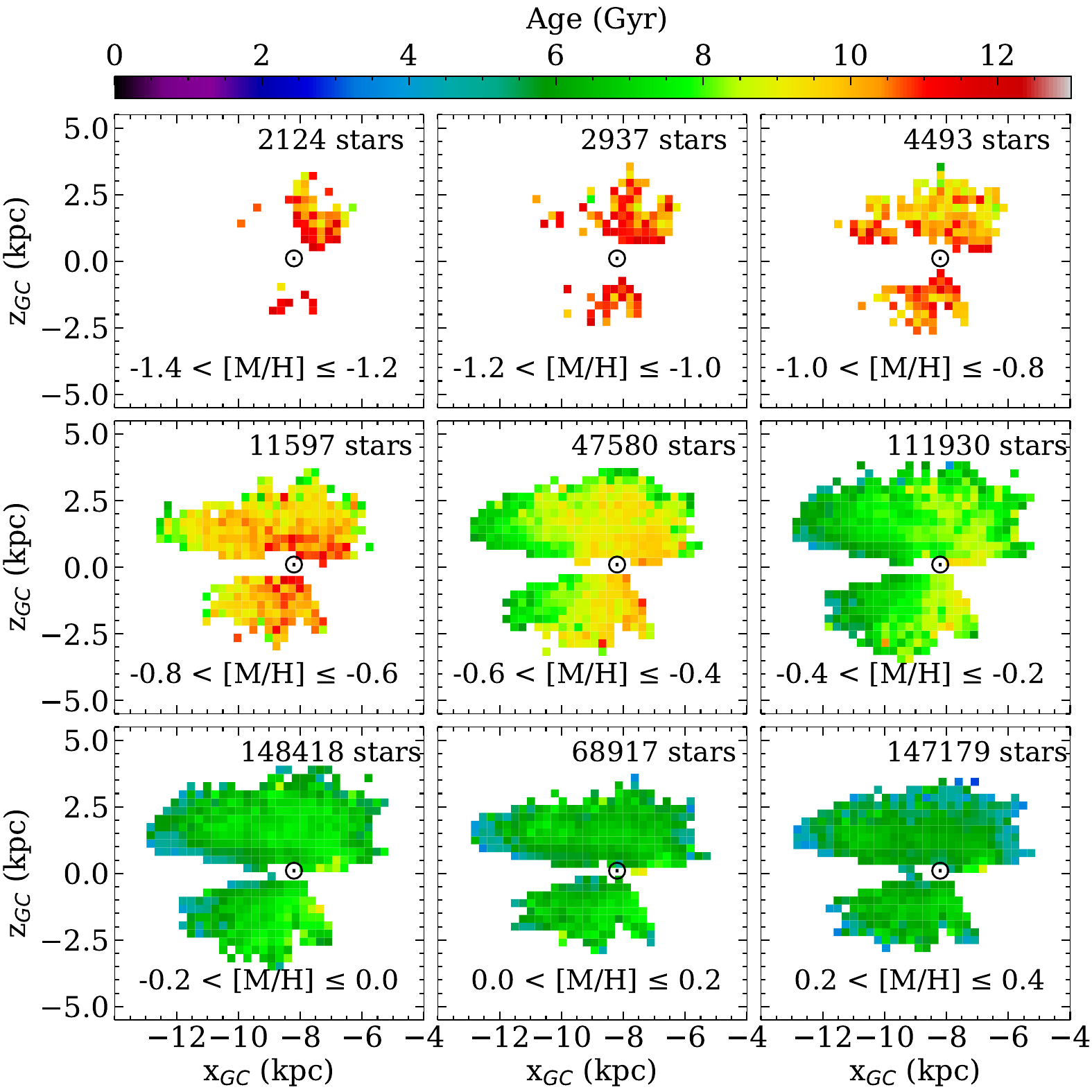}
    \caption{Age map in galactocentric $x_{GC}$ and $y_{GC}$ coordinates of the \jpxgf ($\rsds\leq 4.2$ mag) sample.
    The Age values were derived as described in Sec.~\ref{ssec:zcorr}, using Grid E.
    The colour-bar indicates the mean a for stars within $0.2\times0.2$ kpc pixels.
    The pixels with less than $10$ stars were rejected.
    }
    \label{fig:xz_mean_age}
\end{figure}

We derived individual ages, metallicities, and \afe\ abundances for the \jpxgf\ stars by matching their J-PLUS photometry and \emph{Gaia} parallaxes to the 1152 isochrones of Grid~E.
For every star $n$ and isochrone $i$ we computed the likelihood in Eq.~\ref{eq:bayes} and adopted the parameters of the isochrone that maximised it.
Apparent magnitudes and parallaxes, rather than absolute magnitudes, were fitted, ensuring internal consistency between the inferred parameters and the three–dimensional position of each star.

Fig.~\ref{fig:xz_mean_mh}, Fig.~\ref{fig:xz_mean_age} and Fig.~\ref{fig:xz_mean_afe} display the spatial trends of metallicity, age and \afe, respectively.
Each mosaic stacks nine {\mh} (age) slices, revealing systematic variations with height above the plane and Galactocentric radius.
The corresponding pixel–by–pixel dispersions are presented in Figs.~\ref{fig:3x3_std_mh}, \ref{fig:3x3_std_age} and Fig.~\ref{fig:3x3_std_afe}.
Because they include the physical spread in age and metallicity along the $y_{GC}$ axis, these dispersions exceed the random photometric errors.

\subsubsection{Young and Intermediate-age Populations ($<7$ Gyr)}\label{sssec:youngstar}

In Fig.~\ref{fig:xz_mean_mh}, stars younger than 7 Gyr show metallicities around $0.0 \lesssim \mh < 0.1$~dex within $|z_{\mathrm{GC}}| \lesssim 1$~kpc, increasing to $\mh \gtrsim 0.1$~dex for $|z_{\mathrm{GC}}| \lesssim 0.75$~kpc.
This distribution is similar to the observed in APOGEE DR17 \citep[][their Fig.~5]{imig23}, where \feh$<-0.2$ for $\vert z/{\rm kpc}\vert<1$.
A small fraction of young stars with $\mh\approx0$~dex reaches $|z_{GC}|\simeq1$–2~kpc, and those with $\mh\le-0.1$~dex extend even higher.
Although these heights exceed the canonical thin-disc scale height of $\sim200$~pc~\citep[e.g.][]{bovy17}, similar low-\afe\ stars are observed at $1\lesssim|z|/{\rm kpc}\lesssim2$ and $6<R_{\rm GC}/{\rm kpc}<9$ in the APOGEE sample \citep[Fig. 8 from][]{imig23}.

The mean age distribution for different metallicity bins is shown in Fig.~\ref{fig:xz_mean_age}.
For metal-poor stars ($\mh\leq-0.4$ dex), there is little to no age gradient, consistent with the uniform distribution of old, metal-poor stars from the halo and thick disk.
At intermediate metallicities ($-0.4$ to $0.2$ dex), however, we observe a mild gradient in the mean stellar age with respect to $z_{\rm GC}$.
This is precisely the metallicity range where the two populations identified in Figs.~\ref{fig:amr_basti} and \ref{fig:amr_basti2} coexist, but with different ages.
Young stars ($<6$ Gyr) with $-0.2 < \mh /{\rm dex}\leq 0.2$ dominate the mean age within 1 kpc and toward the Galactic anticentre (cyan and dark-green pixels), whereas older stars (6–8 Gyr) dominate at higher vertical positions ($z_{GC} > 1$ kpc).
Some spurious effects appear in pixels at larger heliocentric distances, especially in the high-metallicity range, where unrealistically young ages are inferred.
Such ages are not consistent with expectations at those vertical positions.
These effects are likely caused by the small number of stars and parallax errors at large distances, since distance affects both the derived age and the Bayestar 3D reddening-map solution.
Indeed, we observe large age variance in these pixels (see Fig.~\ref{fig:3x3_std_age}).
In contrast, metallicity depends mainly on color and is therefore far less sensitive to distance errors.

\subsubsection{Old Populations ($>8$ Gyr)}\label{sssec:oldstar}

For ages older than 8 Gyr, stars with $-0.4 \lesssim \mh/{\rm dex} \lesssim -0.1$~dex dominate the distribution around $|z_{GC}| \simeq 2$~kpc, extending farther than their more metal-rich counterparts ($\mh \gtrsim -0.1$~dex). The oldest and most metal-poor stars ($\mh \lesssim -0.6$~dex) lie even higher and probably belong to the thick disc, with a small fraction originating from the stellar halo.
A pronounced vertical metallicity gradient appears in the $9.5<\mathrm{Age/Gyr}\le12.6$ panel,
dropping from $\mh\simeq-0.1$~dex at the plane to $\mh\lesssim-0.7$~dex at large $|z_{GC}|$.

Fig.~\ref{fig:xz_mean_age} reveals no obvious age–$z_{GC}$ trend for the old sample, but a clear dependence on $x_{GC}$:
the relative number of 8–10.5~Gyr stars increases towards the Galactic centre, consistent with the stellar density profile of the thick disc.
Stars older than 10.5~Gyr dominate at $\mh\lesssim-0.6$~dex and become exclusive below
$\mh\approx-1$~dex, in line with a halo origin.

See Appendix~\ref{app:comp_results} for further information about the distribution of \afe\ as a function of the $x_{GC}$ and $z_{GC}$ coordinates.


\section{Discussion}\label{sec:discus}

The SFH of the \jpxgf shows at least two populations with different chemical evolution. The first one, formed by $\alpha$-enhanced stars born between 12.5 to 8 Gyrs ago.
The second one, which seems to be formed mainly by solar-scaled stars, evolves from $\sim8$ Gyr ago to nowadays. In this Section we discuss both populations and compare our results with other works.

\subsection{Signatures of the Galactic thick disc formation}

The age distribution of \afe=0.4 dex in Fig.~\ref{fig:amr_basti2} reveals a chemical enrichment that began at oldest ages, between 13.7 to 11.3 Gyr ago\footnote{In isochrone grids B,C, D and E, the oldest age is $\approx12.5$ Gyr and covers the interval ranging from 13.7 to 11.3 Gyr.}, and was relatively short.
Approximately, within $\approx4.5$ Gyr the interstellar medium (ISM) was driven from $\mh\approx-0.5$ dex to supersolar values as the sequence ends around $8$ Gyr ago.
This old high-$\alpha$ component of the SFH is likely linked to a significant number of thick disc stars from the \jpxgf same, and can be related to a relatively short an intense formation episode this component.
Independent large–survey studies report a similar chronology.
\citet{sahlholdt22} show how GALAH MSTO stars outline one narrow age–metallicity ridge that terminates about $10$ Gyr ago, implying an equally intense early star-formation episode.
APOGEE mapping of high-$\alpha$ stars finds a median age $\lesssim 9$ Gyr throughout the disc, again suggesting a rapid first phase despite later selection biases \citep{cerqui25}.
The {\g} colour–magnitude analysis made by \citet{alvar25} push the onset even earlier, at $13$ Gyr, and shows enrichment to supersolar metallicity by $10$ Gyr,
while the dynamically selected sample of LAMOST SGB stars in \citet{xiang22} trace a remarkably tight $[{\rm Fe}/{\rm H}]$-age track from $-1$ dex at $13$ Gyr to $0.5$ dex by 7-8 Gyr, implying a well-mixed ISM during the entire high-$\alpha$ phase.

The well shaped enrichment trend we recovered in Fig.~\ref{fig:amr_basti2} for \afe=0.4 dex, even when observational age scatter is present,
matches the expected when a dense and well-mixed gas reservoir is consumed in a brief, intense star-formation episode.
In the simulations of \citet{khopers21} this phase reaches $\geq 10~{\rm M}_{\odot}\cdot{\rm yr}^{-1}$ for several hundred Myr, producing a compact, high-$\alpha$ disc whose chemistry evolves almost synchronously across radius because the ISM is efficiently stirred and enriched.
The Auriga simulation from \citet{grand18} show a centrally concentrated starburst, often merger-triggered, drives the ISM from \feh$\approx -0.5$ dex to supersolar within $\lesssim 3$ Gyr, after which the gas disc contracts, the SFR decreases sharply, and a chemically distinct low-$\alpha$ sequence begins to grow from newly accreted gas.
\citet{agertz21}’s VINTERGATAN simulation link a similar transition to the Galaxy’s last major merger. While the inner high-$\alpha$ disc survives, the encounter seeds an extended, metal-poor and low-$\alpha$ outer disc, whose lower SFR mode soon dominates the posterior generation of stars.
Our SFH shows a similar turning point, marked by a sharp decline in $\alpha$-rich star formation after $\approx 9$ Gyr, accompanied by a rise in solar-scaled star formation from 10 Gyr towards younger ages.

In general, major mergers can significantly affect the internal star formation of the host galaxy.
For instance, simulations by \citet{moreno15} show that the first peri-centric passage of the less massive companion enhances star formation in the central regions of the primary galaxy.
In this context, \citet{alvar25} link their 10.5~Gyr enhancement of star formation explicitly to the GSE merger.
While the uncertainties in our results limit the scope of the discussion, our inferred SFHs remain consistent with a scenario in which the GSE merger contributed to shaping the early disc.
This motivates extending the analysis by incorporating spatial and/or kinematical selections to explore possible hidden features in the SFHs.

The \afe=0.4 SFH in Fig.~\ref{fig:amr_basti2} shows a population of stars with $\mh>0$ at 7-9 Gyr.
Given the strong incompleteness of our \jpxgf ($\rsds\leq4.2$ mag) sample below $\vert z_{\rm GC}\vert< 400$ pc, the number of inner-disc migrant stars is significantly reduced.
We therefore interpret these $\mh > 0$ stars as likely representing the tail end of the early, well-mixed chemical evolution, rather than later arrivals from the inner Galactic region.
\citet{xiang22} report that their high-$\alpha$ sequence naturally extends to $0.5$ dex by $7$ Gyr,
produced in situ from a thoroughly stirred ISM rather than by radial migration.

Collectively, these studies reinforce the view that the Milky Way’s thick disc was forged during a short, intense, central burst, and that the drop in $\alpha$-rich star formation around $9$ Gyr ago signals the switch to the less intense, low-$\alpha$ regime traced by the thin disc.

\subsection{Clues to the Galactic thin disc history}

Our results reveal a chemically distinct stellar population for ages younger than $8$ Gyr, characterized by a low-$\alpha$ abundance pattern and a slower metallicity enrichment compared to the $\alpha$-enhanced population.
This is consistent with the scenario of a different star formation regime, in which the interstellar medium becomes enriched primarily through Type Ia supernovae over longer timescales.
In this context, \citet{alvar25} report a more gradual chemical evolution in the thin-disc component of their kinematically selected sample, with a particularly modest enrichment in the SFH for stars younger than 6 Gyr.
Similarly, the low-$\alpha$ population analysed in \citet{xiang22} displays a significantly slower chemical enrichment than its high-$\alpha$ counterpart.
The intermediate/young sequence traced by the blue dot-dashed line in Fig.~\ref{fig:amr_basti}a are in agreement with these findings, which we identify as a low-$\alpha$ population forming under quieter conditions and in a more chemically evolved environment than the thick disc progenitors.
Interestingly, this sequence coexists in time with the tail end of the $\alpha$-enhanced formation period (see Fig.~\ref{fig:amr_basti}b and Fig.~\ref{fig:amr_basti2}).
This behaviour could reflect a dilution episode in the ISM, as the solar-scaled isochrones better match the substructure below the blue track in Fig.~\ref{fig:amr_basti2} for \afe=0.0 dex, while the $\alpha$-enhanced isochrones remain appropriate above it.

Such dilution patterns are naturally explained by simulations. \citet{khopers21} demonstrate that the strong feedback responsible for halting thick-disc formation can drive metal-rich gas out of the star-forming regions, after which the ISM becomes diluted by inflowing, metal-poor gas from the halo or outer disc.
This resumes star formation at a lower rate and over longer timescales, generating a low-$\alpha$ population in a chemically distinct sequence.
Similarly, the VINTERGATAN simulation by \citet{agertz21} shows that the last major merger leads to a moderate mode of star formation, effectively ending the thick-disc era.
In their scenario, the thin disc gradually builds up from cosmological accretion and gas stripped from satellites, forming stars with lower {\afe} in an extended disc configuration.
Although \citet{grand18} describe a different mechanism, a shrinking and subsequent regrowth of the gas disc from external accretion, the outcome is similar: a chemically separate, low-$\alpha$ sequence emerges following the peak of early star formation.

Additional evidence of radial migration may also be present in our results. In Fig.~\ref{fig:amr_basti2} for \afe= 0.0 dex, we identify stars with $\mh>0.0$ and ages older than $6$ Gyr, which could be interpreted as stars born in the inner disc, after the thick-disc phase, in a metal-rich but low-$\alpha$ environment, and later redistributed outward via secular migration.
This is consistent with the timescales expected for stars to reach larger orbital radii under realistic migration rates.
Finally, we note that incompleteness in our sample near the Galactic plane could hinder the detection of stars born in localized star formation events.
Notably, the starburst episode linked to the passage of the Sagittarius dwarf galaxy, possibly around 5–6 Gyr ago as proposed by \citet{ruizlara20} and \citet{alvar25}, may be underrepresented in our results due to this selection limit.

\section{Conclusions}\label{sec:conclusions}

We inferred the star formation history of a sample of stars located within 5 kpc and the magnitude limits $m_r=14$ mag and $m_r=16.5$ mag.
Our analysis combined trigonometric parallaxes with 12 photometric bands from the {\g} and J-PLUS DR3 catalogues, using a Bayesian CMD-parallax fitting approach.
This method supports any number of filters, incorporates prior distributions on metallicity, accounts for selection effects due to magnitude limits, and fits distance and photometry in a fully consistent framework.

The simultaneous fitting of solar-scaled and $\alpha$-enhanced isochrones reveals distinct formation epochs for chemically different stellar populations in the Galactic disc.
The $\alpha$-enhanced component dominates the early star formation history, with most stars forming between 12.5 and 8 Gyr ago.
This period exhibits a well-defined chemical enrichment sequence, consistent with intense star formation driven by a dense, efficiently mixed interstellar medium.
This result is in agreement with previous observational works like \citep{xiang22, alvar25} and simulations such as those by \citet{khopers21} and \citet{grand18} 

In contrast, the solar-scaled population is dominated by intermediate-age stars formed between 9 and 3 Gyr ago, with only a minor contribution from older stars.
This population exhibits a slower chemical enrichment trend, characteristic of a lower star formation regime, with iron abundances primarily increased through Type Ia supernovae.
A slightly decreasing metallicity trend is also observed at super-solar values between ages 8 and 5 Gyr, suggesting that radial migration contributes to the present-day composition of the thin disc.
The presence of multiple sub-populations, spanning both solar and sub-solar metallicities, indicates a more complex and prolonged star formation history.
These stars likely trace the evolutionary phases of the thin disc, shaped by extended chemical evolution driven by secular processes rather than rapid early enrichment.

The resulting maps of mean $\mh$, age, and \afe{} across the Galactic plane reveal coherent spatial trends.
For old stars ($9.5<\mathrm{Age/Gyr}\leq12.6$), the vertical metallicity gradient is pronounced, decreasing from $\mh \simeq -0.1$dex near the plane to $\mh \lesssim -0.7$dex at large $|z_{GC}|$.
The fraction of stars with ages between 8 and 10.5 Gyr increases toward the inner disc, consistent with a thick-disc density profile, while stars with $\mh \lesssim -1$dex are found to be almost exclusively older than 10.5 Gyr, as expected for a stellar halo component.
In addition, stars younger than 7~Gyr are generally metal-rich and concentrated close to the plane, consistent with the expected properties of the thin disc populations.

The multi-filter photometry from J-PLUS has been instrumental in mitigating the age–metallicity and $\alpha$-element degeneracies inherent to CMD fitting based on broadband photometry with a low number of filters. By incorporating strategically chosen narrow and broad filters sensitive to key stellar absorption features, we achieve a more robust characterisation of stellar metallicities and elemental abundances.
This chemical sensitivity allowed us to derive an age-independent metallicity distribution, which was later incorporated into our analysis as a prior on $\mh$, improving the reliability of the SFH inference across the CAMD.
The combination of J-PLUS photometry, our methodology, and theoretical stellar evolution models with varying $\alpha$-element compositions allows the simultaneous identification of high- and low-$\alpha$ populations, demonstrating the potential of J-PLUS data to constrain distinct star formation regimes and chemical enrichment histories within the Milky Way.
Furthermore, the methodology presented here is well suited to exploit large photometric datasets such as J-PAS, S-PLUS, and LSST, and is versatile enough to incorporate individual stellar metallicities and kinematics, enabling robust and detailed analysis of the Milky Way’s star formation history across spatial, temporal, and chemical scales.

\section{Data availability}



The files containing the inferred $a_i$ values used in Figs.~\ref{fig:amr_parsec}, \ref{fig:amr_basti}, and \ref{fig:amr_basti2}, together with the catalogue containing the individual stellar ages and metallicities used in Figs.~\ref{fig:xz_mean_mh}, \ref{fig:xz_mean_age} and \ref{fig:xz_mean_afe}, are available at Zenodo (\href{https://doi.org/10.5281/zenodo.17592911}{DOI: 10.5281/zenodo.17592911}).

\begin{acknowledgements}

J. A. A. T. acknowledges the financial support from the European Union - NextGenerationEU through the Recovery and Resilience Facility (RRF) program Planes Complementarios con las CCAA de Astrof\'{\i}sica y F\'{\i}sica de Altas Energ\'{\i}as - LA4. A. E. acknowledges the financial support from the Spanish Ministry of Science and Innovation and the European Union - NextGenerationEU through the RRF project ICTS-MRR-2021-03-CEFCA. A. H. acknowledges the financial support of the Spanish Ministry of Science and Innovation (MCIN/AEI/10.13039/501100011033) and FEDER, A way of making Europe with grant PID2021-124918NB-C41

A.~del~Pino acknowledges financial support from the \emph{Severo Ochoa} grant CEX2021-001131-S (MICIU/AEI/10.13039/501100011033), the Ram\'on y Cajal fellowship RYC2022-038448-I (MICIU/AEI/10.13039/501100011033, co-funded by the European Social Fund Plus), and the RyC-MAX grant 20245MAX008 (CSIC).

Based on observations made with the JAST80 telescope and T80Cam camera for the J-PLUS project at the Observatorio Astrof\'{\i}sico de Javalambre (OAJ), in Teruel, owned, managed, and operated by the Centro de Estudios de F\'{\i}sica del Cosmos de Arag\'on (CEFCA).
We acknowledge the OAJ Data Processing and Archiving Unit (UPAD; \citealt{upad}) for reducing the OAJ data used in this work.

Funding for the J-PLUS Project has been provided by the Governments of Spain and Arag\'on through the Fondo de Inversiones de Teruel;
the Aragonese Government through the Research Groups E96, E103, E16\_17R, E16\_20R, and E16\_23R;
the Spanish Ministry of Science and Innovation (MCIN/AEI/10.13039/501100011033 y FEDER, Una manera de hacer Europa) with grants PID2021-124918NB-C41, PID2021-124918NB-C42, PID2021-124918NA-C43, and PID2021-124918NB-C44;
the Spanish Ministry of Science, Innovation and Universities (MCIU/AEI/FEDER, UE) with grants PGC2018-097585-B-C21 and PGC2018-097585-B-C22;
the Spanish Ministry of Economy and Competitiveness (MINECO) under AYA2015-66211-C2-1-P, AYA2015-66211-C2-2, AYA2012-30789, and ICTS-2009-14; and European FEDER funding (FCDD10-4E-867, FCDD13-4E-2685).
The Brazilian agencies FINEP, FAPESP, and the National Observatory of Brazil have also contributed to this Project.

This work presents results from the European Space Agency (ESA) space mission Gaia. Gaia data are being processed by the Gaia Data Processing and Analysis Consortium (DPAC). Funding for the DPAC is provided by national institutions, in particular the institutions participating in the Gaia MultiLateral Agreement (MLA). The Gaia mission website is \url{https://www.cosmos.esa.int/gaia}. The Gaia archive website is \url{https://archives.esac.esa.int/gaia}.

This research has partially funded by MICIU/AEI/10.13039/501100011033/ through grant PID2023-146210NB-I00.

AAC acknowledges financial support from the Severo Ochoa grant CEX2021-001131-S funded by MCIN/AEI/10.13039/501100011033 and the project PID2023-153123NB-I00 funded by MCIN/AEI.

LLN thanks Funda\c{c}\~ao de Amparo \`a Pesquisa do Estado do Rio de Janeiro (FAPERJ) for granting the postdoctoral research fellowship E-40/2021(280692).

The work of V.M.P. is supported by NOIRLab, which is managed by the Association of Universities for Research in Astronomy (AURA) under a cooperative agreement with the U.S. National Science Foundation.

\end{acknowledgements}


\bibliographystyle{aa}
\bibliography{reference}


\appendix

\section{Data query}\label{app:selfun}

To ensure the reliability of our photometric measurements, we selected sources using the following quality criteria: (i) the maximum SExtractor \texttt{FLAGS} value across all filters is less than 3, permitting only minor detection issues (e.g., proximity to bright neighbours or image borders), except for sources in the $m_r$ band where we additionally allow \texttt{FLAGS} values between 2048 and 2051 to include forced photometry cases; (ii) the maximum \texttt{MASK\_FLAGS} value is less than 1, thereby excluding sources affected by masked or defective pixels; and (iii) the minimum normalised weight map value (\texttt{NORM\_WMAP\_VAL}) across all filters exceeds 0.8, ensuring high data quality in all bands. 
These criteria collectively provide a robust sample for our study of the Galactic disc.

We applied the $2\sigma$ confidence interval criteria from \citet{delpino24} to achieve a purer sample of stars by imposing hard constraints on the probabilities of an object belonging to each of the three classified categories: Galaxy, QSO, or Star. Specifically, for an object to be classified as a star at the $2\sigma$ confidence level, the 98th percentile of its probability of being a galaxy (\texttt{PCGalaxy(98)}) and the 98th percentile of its probability of being a QSO (\texttt{PCQSO(98)}) must both be less than 1/3. Simultaneously, the 2nd percentile of its probability of being a star (\texttt{PCStar(02)}) must be greater than 1/3. This approach results in a highly confident stellar sample, though it may lead to a reduction in the sample's overall completeness compared to less stringent criteria

\begin{lstlisting}

SELECT
-- Columns --
MagAB.TILE_ID,
MagAB.NUMBER,
MagAB_F.ALPHA_J2000,
MagAB_F.DELTA_J2000,
MagAB.MAG_APER_COR_3_0,
MagAB_F.MAG_ERR_APER_3_0,

FROM jplus.MagABDualPointSources AS MagAB
-- Join with other catalogs --
LEFT JOIN jplus.MagABDualObj AS MagAB_F ON ((MagAB.NUMBER = MagAB_F.NUMBER) AND (MagAB.TILE_ID = MagAB_F.TILE_ID))
LEFT JOIN jplus.ClassBANNJOS AS BANNJOS ON ((MagAB.NUMBER = BANNJOS.NUMBER) AND (MagAB.TILE_ID = BANNJOS.TILE_ID))

WHERE
-- Good photometry --
(ARRAY_MAX_INT(MagAB_F.FLAGS)<3 OR MagAB_F.FLAGS[jplus::rSDSS] BETWEEN 2048 AND 2051)
AND ARRAY_MAX_INT(MagAB_F.MASK_FLAGS)<1
AND ARRAY_MIN_FLOAT(MagAB_F.NORM_WMAP_VAL)>0.8
-- BANNJOS classification --
AND ( (BANNJOS.CLASS_STAR_prob_pc02 >=  0.333)
AND (BANNJOS.CLASS_QSO_prob_pc98 < 0.333)
AND (BANNJOS.CLASS_GALAXY_prob_pc98 < 0.333) )
AND (MagAB.MAG_APER_COR_3_0[1] BETWEEN 14. AND 19.)
\end{lstlisting}

\subsection{APOGEE sample}\label{app:apo_samp}

To ensure a high-quality and reliable sample from the APOGEE DR17 catalog, we selected stars with $\texttt{SNR} \geq 70$ to guarantee precise stellar parameter and abundance determinations. We required stars to have Simple.
Additionally, we required \texttt{STARFLAG}$=0$, retained only those stars for which \texttt{FE\_H\_FLAG}$=0$ and $\varpi\geq0.2$ mas.


\section{Detailed formulations of our Bayesian model}\label{app:stat_equ}

This appendix presents the detailed mathematical formulations that built the Bayesian inference framework employed in this study. In Appendix~\ref{app:prior_like}, we describe the selection and structure of the prior distributions and likelihood functions, providing an explanation of the assumptions and reasoning behind these choices. Appendix~\ref{app:margpost} then derives the marginal posterior distribution, explicitly incorporating a data completeness function to correct for potential biases in the observed sample. See Table~\ref{tab:statmod} to identify all the variables entering in our model and we mentioned in this appendix.

\subsection{Priors and Likelihood functions}\label{app:prior_like}

In our analysis, we assume that the measurements for parallax and photometric magnitudes are subject to normally distributed errors, which is a reasonable assumption for {\g} and J-PLUS data.
Specifically, each measurement is modeled as an independent sample from a Normal distribution, such that $\varpi\sim\mathcal{N}(\hat{\varpi},e_{\varpi})$ and $m_{J^{k}}\sim\mathcal{N}(\hat{m}_{J^{k}},e_{m_J^{k}})$,
where $\hat{\varpi}$ and $\hat{m}_{J^{k}}$ represent the true underlying values predicted by our model, and $e_{\varpi}$ and $e_{m_J^{k}}$ denote the associated measurement uncertainties. Under this framework, the joint likelihood function for the observed data is given by
\begin{equation}\label{eq:likeli}
    p( d_{n} \vert \beta_{n} ) = \mathcal{N}(\varpi_{n} \vert \hat{\varpi}_{n} e_{\varpi,n}) \cdot \prod_{k=1}^{N_F} \mathcal{N} (m_{J,n}^{k}\vert \hat{m}_{J,n}^{k}e^k_{n}),
\end{equation}

\noindent which determines the \say{probability} of observing the data $\varpi_{n},~m_{J,n}^{k}$ given the model parameters.
We introduced the completeness function $S$ separately for clarity, but it directly multiplies the likelihood as $S(\bm{d})\; p(d_{n} \vert \beta_{n} )$. This factor accounts for the probability of detecting a source with given parallax and apparent magnitude.
Although we treat the completeness function $S(\bm{d})$ as a separate parameter for clarity, mathematically it acts as a likelihood factor, making the resultant likelihood $S(\bm{d})\; p(d_{n} \vert \beta_{n} )$ more adequate for our stellar sample. This accounts for the fact that measuring a parallax and magnitude requires first detecting the source, which is precisely what $S$ represents.
Specifically,
\begin{equation}\label{eq:compl}
S(m) \approx 
\begin{cases}
1 & {\rm if }\ \ 14 \leq m_{r}({\rm  mag}) \leq 16.5\ \ {\rm and}\ \ \varpi>0.2\ {\rm mas} \\
0 & {\rm otherwise}
\end{cases}    
\end{equation}

For every isochrone, we choose the {\it exponentially decreasing volume} density prior to predict the distances, that is $p(\rho)\propto \rho^{2}\exp{(-\rho/L)}$.
\cite{bailer15} assessed the efficiency of this prior on the inference of distances from parallaxes. 
Stellar isochrones predicts the position in the CAMD of stars of any metallicity, mass and age.
We convolved the predicted absolute magnitude of each isochrone point at a certain mass $\mathcal{J}^{k}(M)$ with a normal distribution, such that the theoretical absolute magnitude of every source $J^{k}_{n}$ is distributed as $J^{k}_{n}\sim \mathcal{N}(\mathcal{J}^{k}_{i}(M),\sigma^{k}_{i})$, where the standard deviation $\sigma^{k}_{i}$ is a fixed quantity called {\it tolerance} (Table~\ref{tab:statmod}).
In this way, for a set of isochrones, the prior distribution of the random variables $\rho$ and $J^{k}$ given $a_i$ is proposed as the following mathematical combination
\begin{equation}\label{eq:prior}
     p(r_{n}, J^{k}_{n}\, \vert\, \bm{a}) = \sum_{i=1}^{N_{iso}}\ p_i(\rho_n)\ a_i \int_{M_{l,i}}^{M_{u,i}}\phi(M) \prod_{k=1}^{7}\ \mathcal{N}(J^k_n \vert \mathcal{J}^k_i \sigma_i^k)\ dM.
\end{equation}

We adopt a symmetric Dirichlet distribution as our prior $p(\bm{a})$, which is defined by
\begin{equation}\label{eq:dirichlet}
    P(\bm{a})=\frac{\Gamma(\xi N_{iso})}{\Gamma(\xi)^{N_{iso}}}\prod_{i=1}^{N_{iso}}a_{i}^{\xi-1},
\end{equation}

Notably, when $\xi=1$ this prior becomes uniform, meaning that no single $a_i$ is preferentially weighted. For a more detailed discussion of the Dirichlet distribution, refer to Sec. 3.4 of \citet{gelman13}, and see Sec. 3.2 of \citet{Walms13} for an illustrative example in an astronomical setting.


\subsection{Derivation of the marginal posterior under data completeness}\label{app:margpost}

The marginal posterior distribution from Eq.~\ref{eq:bayes} is computed using Eq.~\ref{eq:likeli}, \ref{eq:prior} and \ref{eq:dirichlet}, and integrating it under the completeness limits from \ref{eq:compl} for stars within 5 kpc, that is

\begin{eqnarray}\label{eq:bayes2}
    p(\bm{a} \vert {\bm d}) &=& p(\bm{a}) \prod_{n=1}^{N_{D}} \int \ \frac{S(d_{n})P(d_{n} \vert \beta_{n}) \ P(\beta_{n}\vert \bm{a},\phi)}{\ell(\bm{a},S)} d\beta_{n}\nonumber\\
    &=& p(\bm{a}) \prod_{n=1}^{N_{D}} \int_{0.0{\rm pc}}^{5000{\rm pc}} d\rho_{n}\ \times\nonumber\\
    &\ &\ \ \ \ \times\ \prod_{k=1}^{N_{F}} \int_{m_{J{\rm B}}^{k}-\dm(\rho_n)}^{m_{J{\rm F}}^{k}-\dm(\rho_n)} P(\bm{a}, \rho_{n}, J_{n}^{k} \vert d_n) \cdot dJ_{n}^{k} \nonumber\\
    &=& p(\bm{a}) \prod_{n=1}^{N_{D}} \frac{\sum_{i=1}^{N_{\rm Iso}} P_{n,i} \cdot a_{i}}{\sum_{i=1}^{N_{\rm Iso}} C_{n,i} \cdot a_{i}} \nonumber\\
    &=& p(\bm{a}) \prod \frac{\bm{P}\cdot\bm{a}}{\bm{C}\cdot\bm{a}},
\end{eqnarray}

\noindent where $m_{J{\rm B}}^{k}-\dm(\rho_n)$ and $m_{J{\rm F}}^{k}-\dm(\rho_n)$ are the bright and the faint limits of $J^{k}_{n}$, derived from the selection function $S(d)$.
Note that the $\ell(S,\bm{a})$ factor is included as $\ell(S,\bm{a})=\sum_{i=1}^{N_{\rm Iso}} C_{n,i} \cdot a_{i}$.
The elements of the $\bm{P}_{N_D\times N_I}$ matrix are given by

\begin{align}\label{eq:pni}
    P_{n,i} & = \int_{r_o}^{r_{lim}} \int_{M_{\rm l}, i}^{M_{\rm u}, i} \mathcal{N}( \varpi_{n}\vert \hat{\varpi}_{n} \sigma_{\varpi, n})\ P(\rho_{n})\ \phi(M)\times \nonumber\\
    &\times \left[ \Phi_{n,i}^{3}(J_{\rm F}) - \Phi_{n,i}^{3}(J_{\rm B}) \right] \prod_{k=1}^{N_F}N_{n,i}^{k}(\rho_{n},M)\ \ dM\ d\rho_{n},
\end{align}
\noindent where
\begin{equation}\label{eq:Nnik}
    N_{n,i}^{k}=\mathcal{N}\left(G_{n}^{k}\Big\vert \mathcal{J}_{i}^{k}+\dm(\rho_n)\ ,\ \sqrt{(e_{n}^{k})^2+(\sigma_{\rm{i}}^{k})^2}\right)
\end{equation}

\begin{align}\label{eq:phi}
    &\Phi_{n,i}^{3}(J_{\rm lim})=\nonumber\\
    &=\Phi\left(-\frac{(J_{n}^{3}-\mathcal{J}_{i}^{3}-\dm_n) -(J_{\rm lim}-\mathcal{J}_{i}^{3}-\dm_n)[1+(e_{n}^{3}/\sigma_{i}^{3})^2] }{e_{n}^{3}\sqrt{1+(e_{n}^{3}/\sigma_{i}^{k})^2}}\right).
\end{align}
\noindent

\ \ \ 
And for the $\textbf{C}_{N_D\times N_I}$ matrix are given by

\begin{align}\label{eq:cni}
    C_{n,i} &= \int_{r_o}^{r_{\rm lim}} \int_{M_{\rm l}, i}^{M_{\rm u}, i} \ dM\ d\rho_{n}\ p(r_{j})\ \phi(M)\ \times \nonumber\\
    &\times \prod_{k=1}^{3}\left[
    \Phi\left(\frac{J_{\rm F}-\left(\mathcal{J}^{k}_{i}+\dm_n \right)}{\sigma^{k}_{i}}\right)-
    \Phi\left(\frac{J_{\rm B}-\left(\mathcal{J}^{k}_{i}+\dm_n \right)}{\sigma^{k}_{i}}\right)
    \right].
\end{align}

In Eq.~\ref{eq:phi} and Eq.~\ref{eq:cni}, $\Phi$ is the cumulative distribution function for a normal probability distribution.


\section{Validation with Mock Catalogs}\label{app:mock_test}

In order to assess the performance and reliability of our modeling approach, we employ mock catalogs that simulate realistic stellar populations under controlled conditions.
This section outlines the methodology used to generate these synthetic datasets and evaluates the accuracy of parameter recovery in various scenarios.
Through these validations, we demonstrate the robustness of our approach and its effectiveness in disentangling complex stellar population characteristics.

\subsection{Simulating Stellar Populations}\label{app:simupop}

Summarizing the large and complex process behind the stellar population synthesis is beyond the scope of this work. See \citet{bressan12} for more information. We just shortly comment here that:

\begin{itemize}
    \item[A)] PARSEC provides the evolutionary tracks in the HR diagram for a particular recipe of physical properties, that is, the code predicts the effective temperature (T$_{\rm eff}$/K) and the luminosity (L/L$\odot$) as function of time for a theoretical star with assigned mass and metallicity values. This process is performed for a range of  mass and metallicity values (see Sec.~\ref{sec:stevol}).
    \item[B)] Posteriorly, another algorithm built isochrones interpolating points of the same age between HR diagram tracks with the same metallicity, drawing a trajectory where all the stars have the same age and metallicity but their masses range from the lower and upper mass limits.
    \item[C)] The code replicates simple stellar populations populating an isochrone by randomly sampling the mass of individual stars from an input mass distribution. For our mock catalogs we selected the Kroupa initial mass function \citet{kroupa01}.
    \item[D)] The algorithm matches the evolutionary track points to a spectrum taken from a stellar atmosphere library. Then, provided a set of filters and zero-point anchor, the code computes the photometry of all stars in the population.
\end{itemize}

Finally, if a stellar population born instantaneously, preserving the same chemical abundances and accumulating a total mass $M_{T}$, PARSEC build mock catalogs transforming this amount of mass to individual stars following the process  summarized in steps A to D. 

From here, we produce our mock catalogs following the following steps:

\begin{enumerate}
    \item We built a source mock catalog with uniform star formation by executing PARSEC for a grid in the intervals ${\rm age}=[2,12]$ Gyr and $\mh=[-0.7,0.2]$ dex, with steps 0.05 Gyr and 0.05 dex respectively. Each point in the grid is a burst of star formation that yields $10^3$ solar masses.
    However, we can increase this mass parameter for specific combinations of age and $\mh$, depending on the case under study. 
    The mass of each of these sub populations will decrease below this value depending on their age and metallicity values. The catalog includes the photometry of the twelve J-PLUS filters.
    \item We assigned heliocentric distances ($\rho$) to all the stars in the source catalog by random sampling the function $p(\rho)\propto (\rho/pc)^{2}\exp(-\rho/pc)$.
    \item We computed the parallax and the twelve apparent magnitudes of each star using the distances from step 2.
    \item We simulated measurements adding normal distributed observational errors on the parallax and magnitudes from step 3.
\end{enumerate}

In this case, we did not aim to build a Milky Way model, but just to build a mock catalog with a CMD similar to the JxG5k sample. For this reason, we did not use a more complex spatial distribution that depends on the galactic coordinates (l,b) in step 2, nor include a relation between age and position. The distances were sampled from the $p(\rho)$ profile using the \texttt{emcee} parackage \citep{foreman13}.

The normal noise in step 3 were added sampling the errors from the \jpxgf sample. Specifically, we created the functions $E_{J^{k}}(\rsds)$ and $E_{\varpi}(\rsds)$ that made groups of stars in our sample with bins of apparent $\rsds$ magnitude and used the parallax and photometric error distribution of each group to assign the corresponding errors to mock stars according to their synthetic apparent $\rsds$ magnitude.

\subsection{Age–Metallicity recovery and the role of data errors}

\begin{figure}
  \centering
  \includegraphics[width=0.85\linewidth]{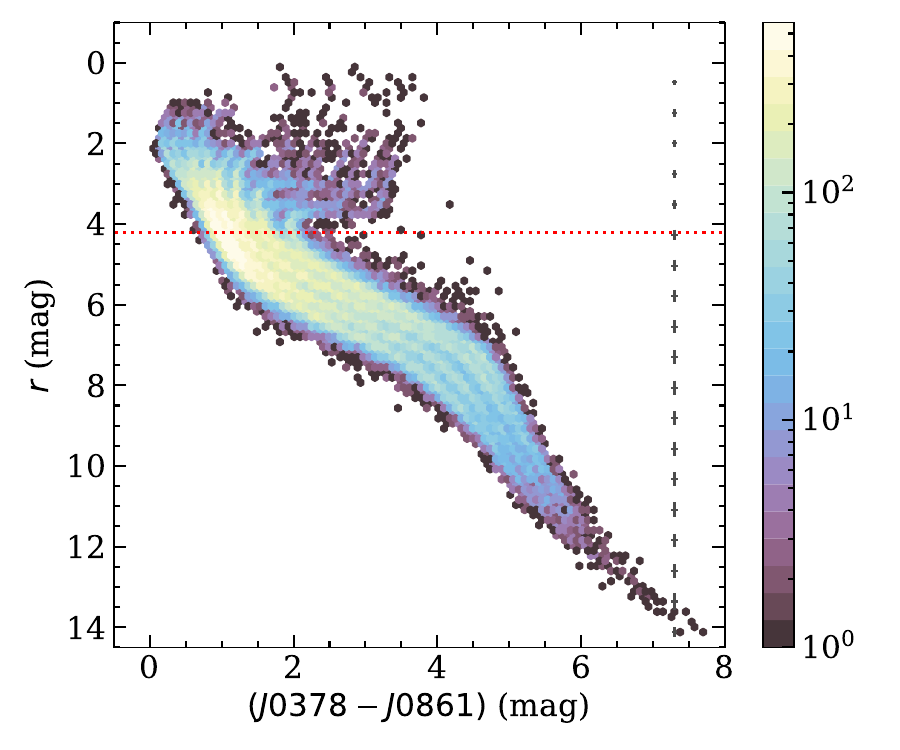}
  \caption{colour-magnitude diagram of the $4\times4$ mock catalog.}
  \label{fig:CMD_mocks}
\end{figure}

We assessed our model inferring the age and metallicities distributions of synthetic stellar populations generated from the source mock catalog (see item 1 from App.~\ref{app:simupop}).
We focused our recovery test on the analysis of a grid of $4\times4$ instantaneous bursts of star formation, with $\mh=-0.8, -0.5, -0.2,\ 0.1$ dex and ages of 1.25, 2.5, 5.0 and 10 Gyr.
Each burst produce $10^{5}\ \msun$ in stars.
We did this test looking to observe the age and metallicity distribution produced by data errors and age-metallicity degeneracy.
Fig.~\ref{fig:CMD_mocks} shows the CAMD of this mock catalog.

We inferred the ages and metallicities of this mock catalog following the same methodology we used to obtain the star formation histories shown in Figs.~\ref{fig:amr_parsec}, \ref{fig:amr_basti}, \ref{fig:amr_basti2}.
First, we derive the burst's metallicities fitting the isochrones of Grid A to stars in the CAMD fainter than $r=4.2$ mag.
In this case, ages are completely biased, but the metallicity distribution is robust.

\begin{figure}
    \centering
    \includegraphics[width=0.9\linewidth]{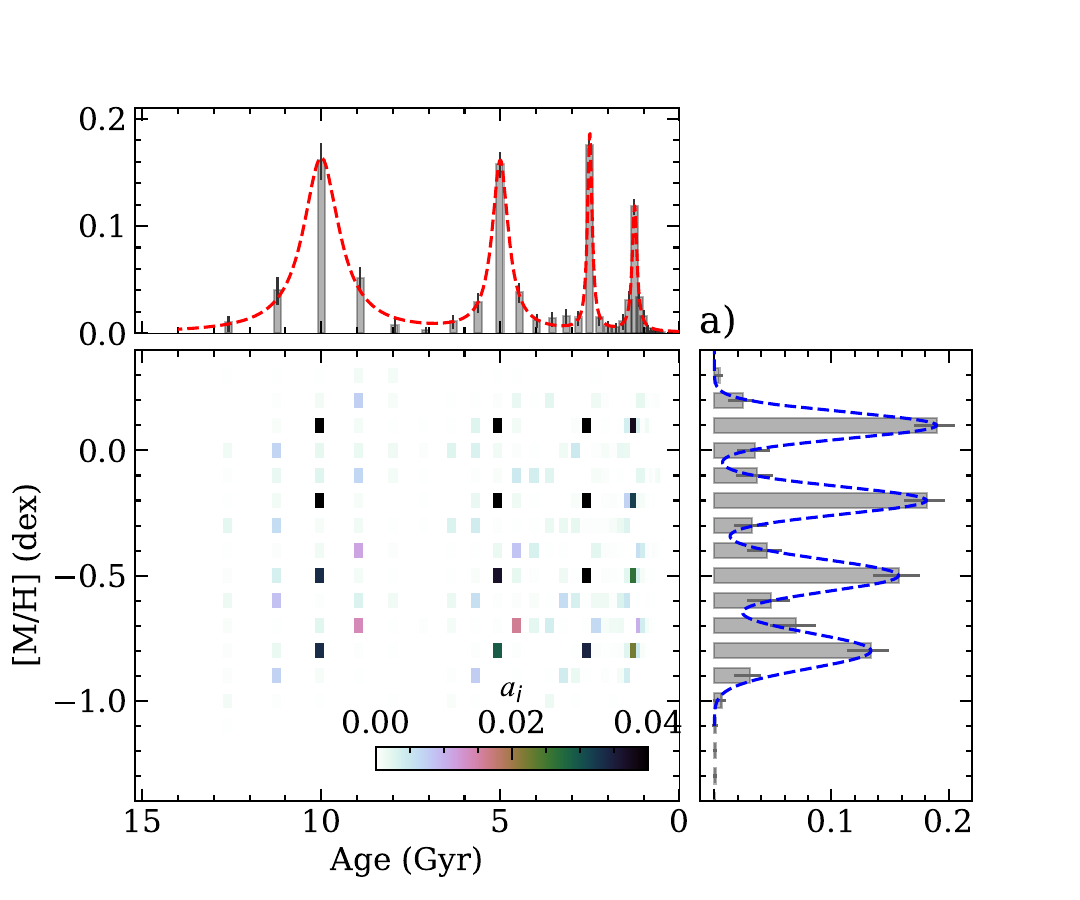}
    \includegraphics[width=0.9\linewidth]{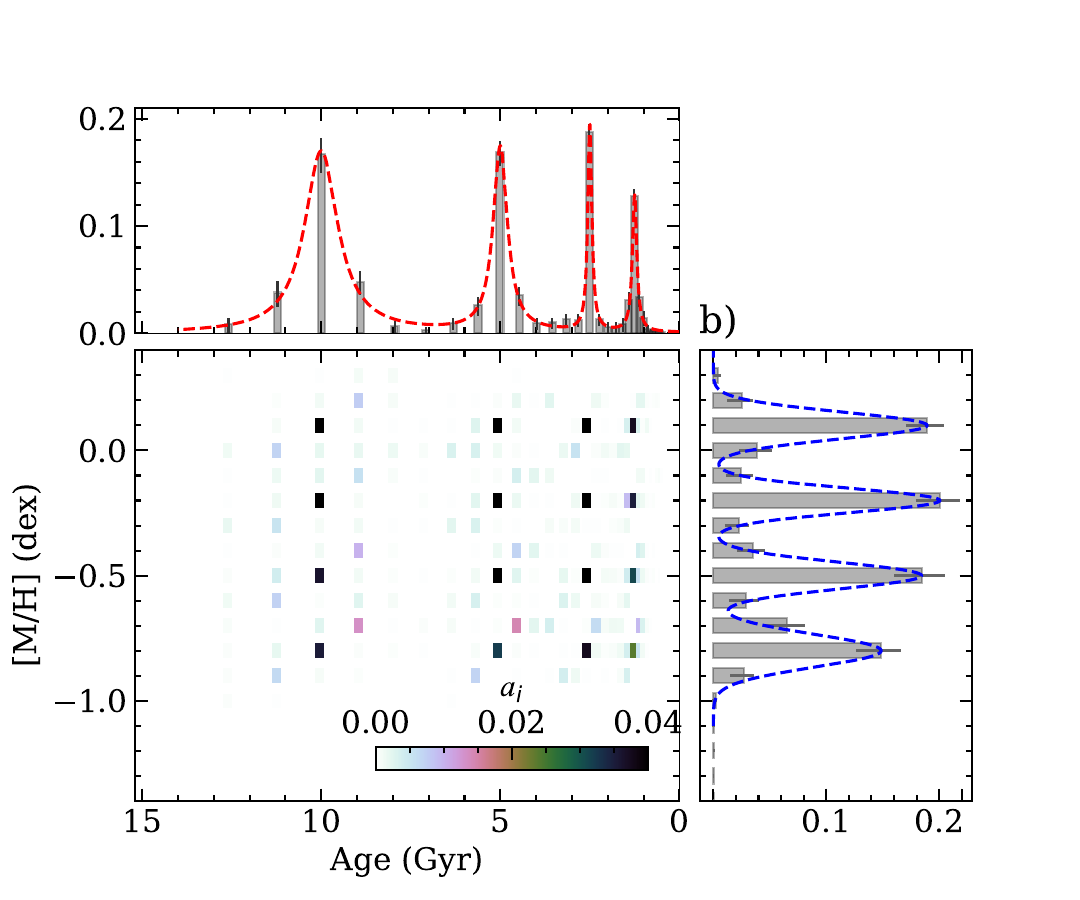}
    \caption{Input and recovered ages and metallicities for the $4 \times 4$ mock catalogue: (a) without and (b) with a prior distribution on $\mh$. The blue dashed lines represent the fit of a model composed of four normal distributions, while the red dashed lines show the fit using a combination of four Cauchy distributions.}
    \label{fig:sfh_mock11}
\end{figure}

Second, we inferred the burst's ages and metallicities by fitting the isochrones of Grid B to stars with $r>4.2$ mag, separating two cases; not using (Fig.~\ref{fig:sfh_mock11}a) and using (Fig.~\ref{fig:sfh_mock11}b) the previous result as a prior on $\mh$.
The blue dashed lines shows the fit of a function composed of four normal distribution functions to the marginal distribution of $\mh$, with different weights, means and standard deviations.
The red dashed lines shows the fit of a function composed of four Cauchy distributions to the marginal distribution of age. 

Table~\ref{tab:fwhm} shows the standard deviation and full width at half maximum (FWHM) of metallicity and age respectively, for the fits shown in Fig.~\ref{fig:CMD_mocks}.
It allows to contrast how much the dispersion is reduced when we use a metallicity prior.
Evidencing how we mitigates the effect of age-metallicity degeneracy on age and metallicity distribution.

We can identify that the burst with lowest metallicity ($\mh$ $=-0.8$ dex) present a null reduction on $2\sigma$, while the larger metallicities show a significant reduction. It is also noticeable how the bias reduction is increasing from young to old ages. This can be noticed on Fig.~\ref{fig:CMD_mocks}.

\begin{table}[ht]
\renewcommand{\arraystretch}{1.25} 
\begin{center}
 \caption{\label{tab:fwhm} Standard deviation ($\sigma$) and FWHM of the Normal and Cauchy distributions, respectively, shown in Fig.~\ref{app:mock_test}. The "No" and "yes" rows indicate the values obtained using and not-using a prior on $\mh$, respectively. The bottom row shows the relative reduction on $\sigma$ and FWHM.}
 \resizebox{9cm}{!}{
 \begin{tabular}{cccccccccc}
 \hline
 \multirow{2}{*}{Prior}   &  \multicolumn{4}{c}{$2\sigma\mh$ (dex)}  &   &  \multicolumn{4}{c}{FWHM age (Gyr)}   \\
                          &     -0.8 & -0.5 & -0.2 & 0.1      &   &       1.25 & 2.5 & 5 & 10       \\ \cline{2-5} \cline{7-10}
  No   & 0.14  & 0.13   & 0.11  & 0.10  &   & 0.16   & 0.15   & 0.55   & 1.22   \\
  Yes  & 0.13  & 0.11   & 0.10  & 0.10  &   & 0.11   & 0.13   & 0.47   & 1.13   \\ \cline{2-5} \cline{7-10}
  $\downarrow$  & 7.1\% & 15.4\% & 9.1\% & 0.0\% &   & 31.3\% & 13.3\% & 14.6\% & 7.5\%  \\
 \hline
 \end{tabular}
 }
\end{center}
\end{table}

\section{Complementary results}\label{app:comp_results}

\begin{figure}
    \centering
    \includegraphics[width=1\linewidth]{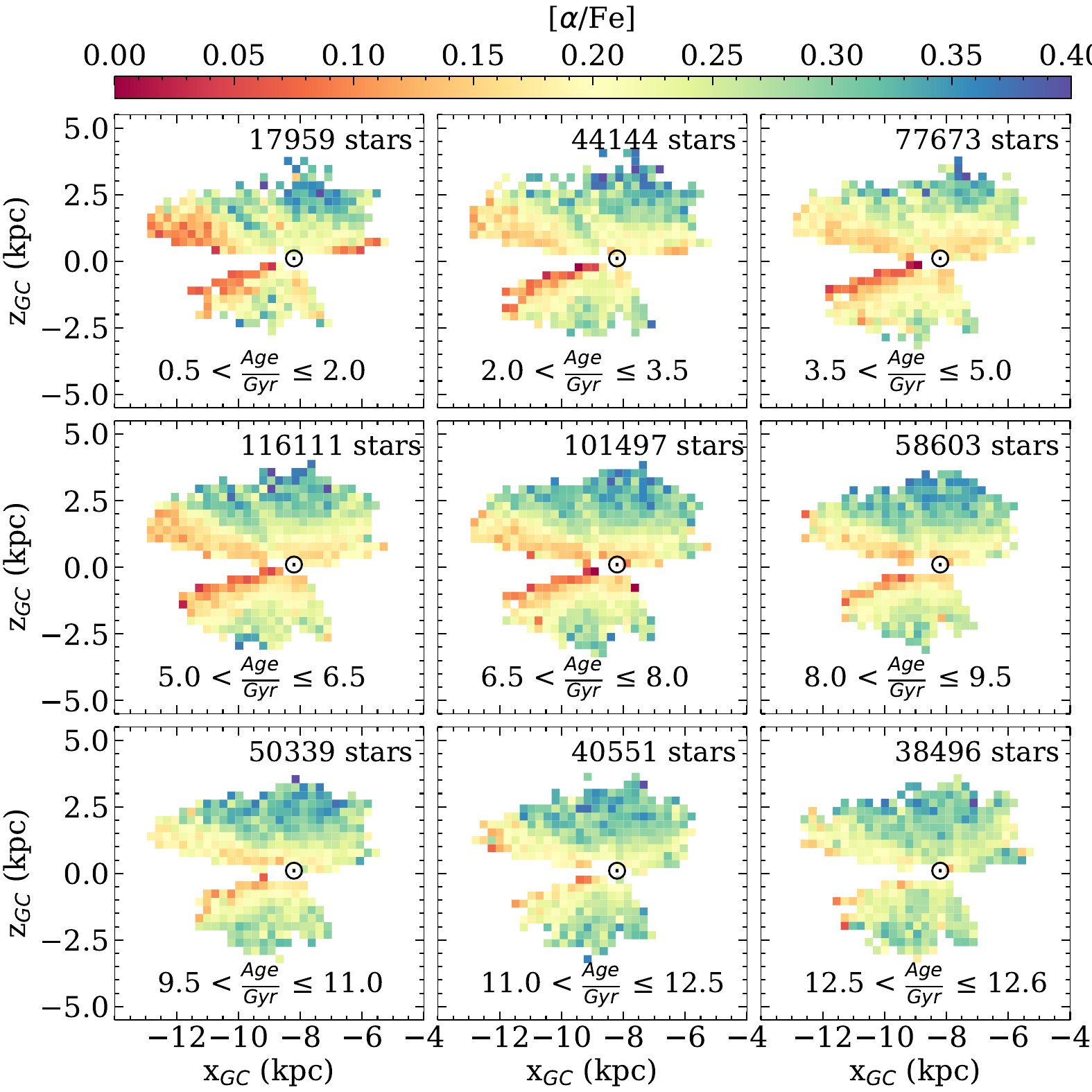}
    \caption{$\alpha$-abundance map in galactocentric $x_{GC}$ and $z_{GC}$ coordinates of the \jpxgf ($\rsds\leq 4.2$ mag) sample.
    The \afe values were derived as described in Sec.~\ref{ssec:zcorr}, using Grid E.
    The colour-bar indicates the mean of \afe for stars within $0.2\times0.2$ kpc$^2$ pixels.
    The pixels with less than 10 stars were rejected.}
    \label{fig:xz_mean_afe}
\end{figure}

Figure~\ref{fig:xz_mean_afe} shows the distribution of \afe\ derived using GridE. These isochrones have only two $\alpha$ compositions;
therefore, the mean \afe\ within each $0.2\times0.2$kpc$^2$ pixel is determined by the relative number of stars with \afe\ = 0.0 or 0.4 dex.
For this reason, a range of colours is visible in Fig.\ref{fig:xz_mean_afe}.
Low-$\alpha$ values (red pixels) are found closer to the Galactic plane and are concentrated towards the Galactic anticentre.
This is consistent with the fact that the number of high-$\alpha$ stars, associated with the inner thick disc, drops significantly beyond the Sun’s position.
High-$\alpha$ values (blue pixels) are found at larger vertical distances from the Galactic plane, consistent with the typical distribution of $\alpha$-abundances in the Milky Way disc \citep{imig23}.
Furthermore, high-$\alpha$ values dominate the pixel distribution for ages older than 10~Gyr.

We shows in this appendix the standard deviation the metallicity age maps presented in Fig~\ref{fig:xz_mean_mh} and Fig~\ref{fig:xz_mean_age}. 

\begin{figure}
    \centering
    \includegraphics[width=1.0\linewidth]{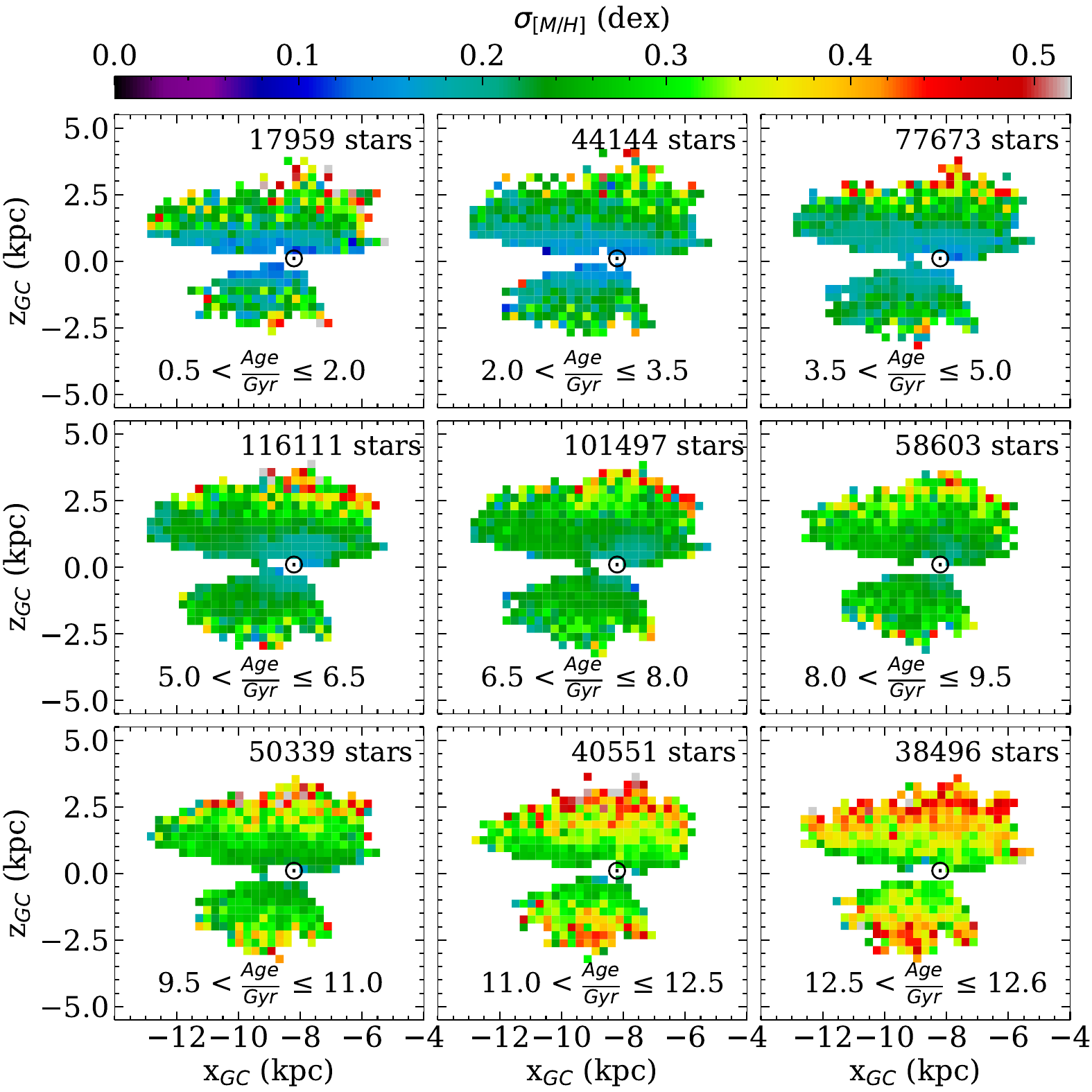}
    \caption{Standard deviation corresponding to the metallicity map in Fig~\ref{fig:xz_mean_mh}.}
    \label{fig:3x3_std_mh}
\end{figure}

\begin{figure}
    \centering
    \includegraphics[width=1.0\linewidth]{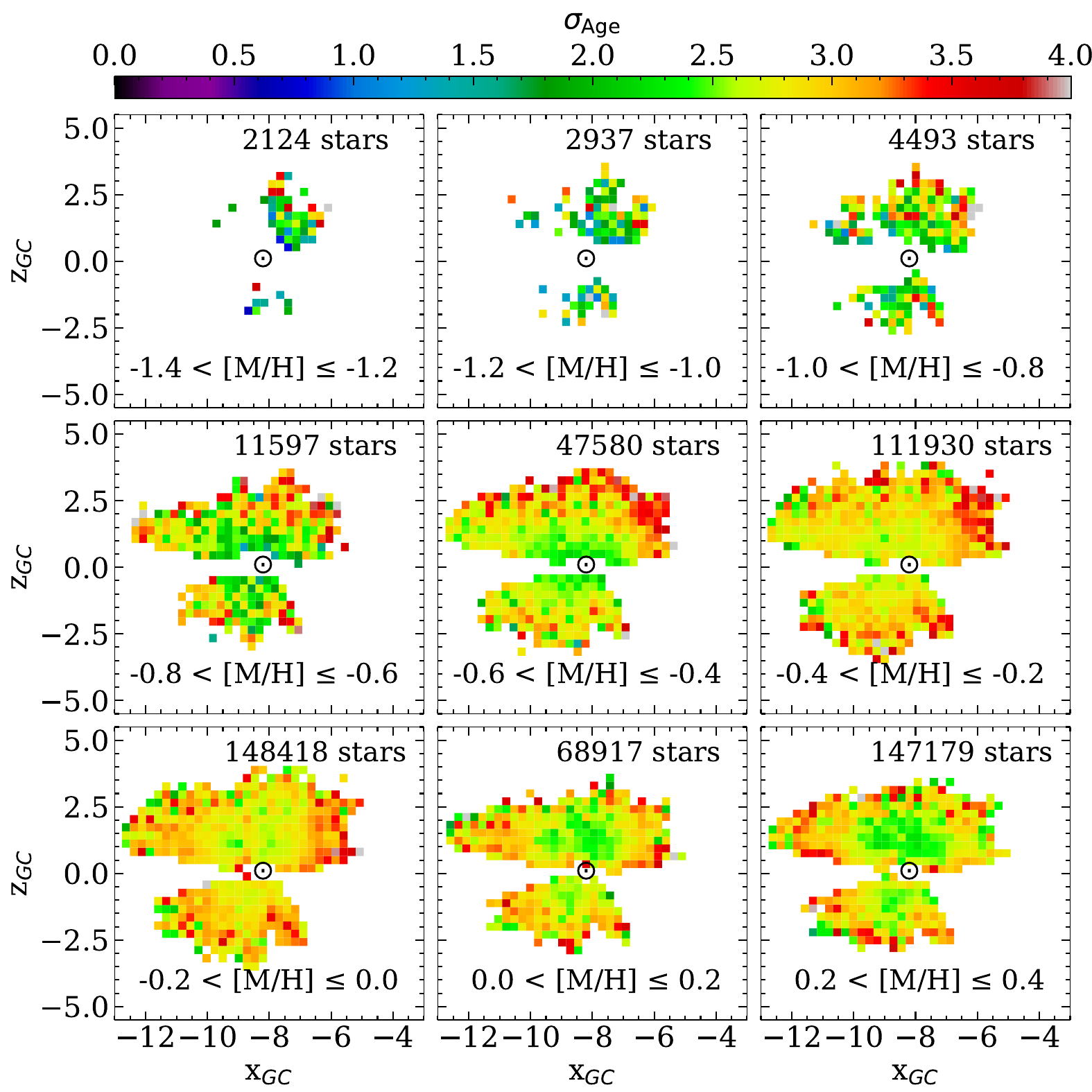}
    \caption{Standard deviation corresponding to the age map in Fig~\ref{fig:xz_mean_age}.}
    \label{fig:3x3_std_age}
\end{figure}

\begin{figure}
    \centering
    \includegraphics[width=1.0\linewidth]{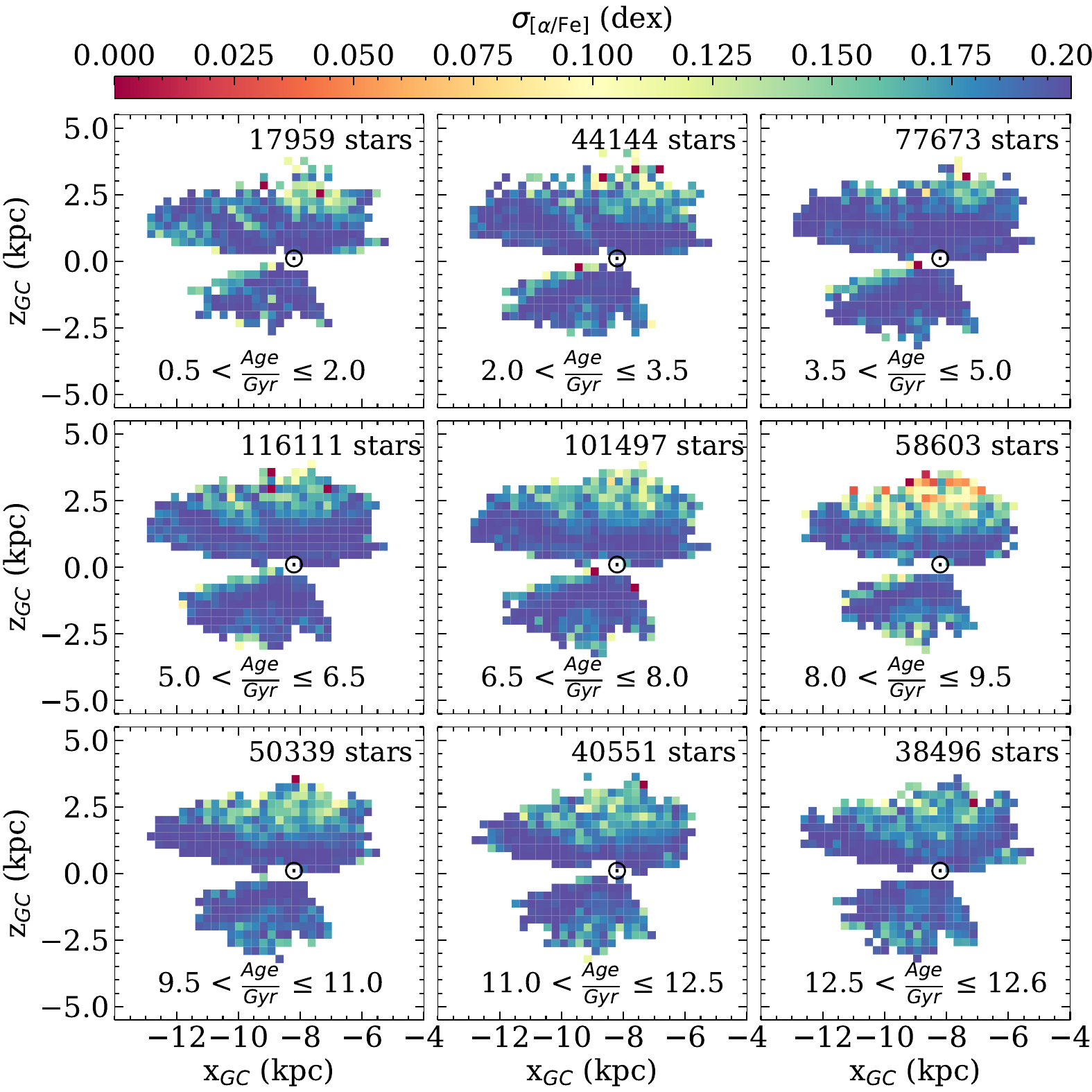}
    \caption{Standard deviation corresponding to the {\afe} map in Fig~\ref{fig:xz_mean_afe}.}
    \label{fig:3x3_std_afe}
\end{figure}

\end{document}